\documentclass[reqno]{article}
\usepackage{amssymb,amsmath,cite,delarray,epsfig,hhline}
\newcommand{\lsim} {\buildrel < \over {_\sim}}
\newcommand{\gsim} {\buildrel > \over {_\sim}}

\begin{document}
\begin{flushright}
\begin{tabular}{l}
UMD-PP-05-023\\OSU-HEP-04-10
\end{tabular}
\end{flushright}

\begin{center}
{\LARGE \bf Tying in CP and Flavor Violations  with \\[0.1in] Fermion
Masses and Neutrino Oscillations }
\vskip.3in ${\large\bf
K.S. ~Babu^a,\ Jogesh\ C.\ Pati^b,\ Parul\ Rastogi^b}$\\[0.1in]
$^a Oklahoma\ Center\ for\ High\ Energy\ Physics,\ Department\ of\
Physics$ \\[-0.03in]
$Oklahoma\ State\ University,\ Stillwater,\ OK\ 74078,\
USA$\\[0.1in]
${\large ^b Department\ of\ Physics,\ University\ of\ Maryland,\ College\ Park\  MD\ 20742,\ USA}$\\

October 13, 2004.
\end{center}

\vskip.25in
\begin{abstract}
In this paper we explore the possibility that (a) fermion masses,
(b) neutrino oscillations,  (c) CP non-conservation and (d) flavor
violations get intimately linked to each other within
supersymmetric grand unification  based on SO(10) or an effective
G(224) = $SU(2)_L\times SU(2)_R\times SU(4)^c$ symmetry.  We
extend the framework proposed previously by Babu, Pati and Wilczek
(BPW) which successfully described fermion masses and neutrino
oscillations, to include CP violation. Assuming flavor universal
SUSY breaking parameters at $M^* \gsim M_{GUT}$, and that CP
violation arises through phases in the fermion mass matrices, we
show how the presence of GUT threshold induces new and calculable
CP and flavor violations. Including SM and SUSY contributions, we
find that the BPW framework can correctly account for the observed
flavor and/or CP violations in  $\Delta m_K,\ \Delta m_{B_d},\
S(B_d\rightarrow J/\psi K_S)$ {\it and} $\epsilon_K$. While
SUSY-contribution is small ($\lsim $ few\%) for the first three
quantities, that to $\epsilon_K$ is sizable ($\sim$ 20-25\%) and
{\it negative} (as desired) compared to that of the SM. The model
predicts $S(B_d\rightarrow \phi K_S)$ to be in the range
+(0.65-0.73), close to the SM prediction. The model yields
$Re(\epsilon'/\epsilon)_{SUSY}\approx +(4-14)\times 10^{-4}$; the
relevance of this contribution can be assessed only when the
associated matrix elements are known reliably. The model also
predicts that the electric dipole moments of the neutron and the
electron should be discovered with improvements in current limits
by factors of 10 to 100.

\end{abstract}
\newpage
{\large

\section{Introduction}
Ongoing studies of CP and flavor violations may well turn out to
provide some important clues to physics beyond the standard model
(SM). Strong agreement with the expectations of the SM would
provide severe constraints on new physics, while departures could
suggest the nature and the threshold for such new physics. On the
experimental side there are now four well measured quantities
reflecting CP and/or $\Delta F = 2$ flavor violations. They
are:$\footnote{$\epsilon'_K$ reflecting direct $\Delta F = 1$ CP
violation is well measured, but its theoretical implications are
at present unclear due to uncertainties in the matrix element. We
discuss this later. }$
\begin{eqnarray}
\label{eq:4qtty} \Delta m_K,\  \epsilon_K,\  \Delta m_{B_d}\ {\rm
and} \ S(B_d \rightarrow J/\Psi K_S)
\end{eqnarray}
where $S(B_d \rightarrow J/\Psi K_S)$ denotes the asymmetry
parameter in ($B_d$ versus $\overline{{B_d}})\rightarrow J/\Psi
K_S$ decays. It is indeed remarkable that the observed values
including the signs of all four quantities as well as the
empirical lower limit on $\Delta m_{B_s}$ can consistently be
realized within the standard CKM model for a single choice of the
Wolfenstein parameters \cite{Ciuchinietal}:
\begin{eqnarray}
\label{eq:rhoeta} \bar{\rho}_W\ = 0.178\ \pm\ 0.046;\
\bar{\eta}_W\ = 0.341\ \pm\ 0.028~.
\end{eqnarray}
This fit is obtained using the observed values of $\epsilon_K$ =
2.27$\times10^{-3}$, $V_{us}$ = 0.2240 $\pm$ 0.0036, $|V_{ub}|$ =
(3.30 $\pm$ 0.24)$\times 10^{-3}$, $|V_{cb}|$
 = (4.14 $\pm$
0.07)$\times 10^{-2}$ , $| \Delta m_{B_d}|$ = (3.3 $\pm$ 0.06)
$\times 10^{-13}$ GeV and $\Delta m_{B_d}/\Delta m_{B_s}$ $
>$ 0.035, and allowing
for uncertainties in the hadronic matrix elements of up to 15\%.
One can then predict the asymmetry parameter $S(B_d \rightarrow
J/\Psi K_S)$ in the SM to be $\approx 0.685 \pm 0.052
$\cite{Ciuchinietal, ASoni}. This agrees remarkably well with the
observed value $S(B_d \rightarrow J/\Psi K_S)_{expt.}$ = $0.734
\pm 0.054$, representing an average of the BABAR and BELLE results
\cite{BabarBelle}. This agreement of the SM prediction with the
data in turn poses a challenge for physics beyond the SM,
especially for supersymmetric grand unified (SUSY GUT) models, as
these generically possess new sources of CP and flavor violations
beyond those of the SM.

The purpose of this paper is to address the issues of CP and
flavor violations $\it{in\  conjunction\  with}$ those of fermion
masses and neutrino oscillations, in the context of SUSY grand
unification. In particular, our goal would be to obtain a
$\it{unified\ description}$, in accord with observations, of all
four phenomena: (i) CP non-conservation, (ii) flavor violation,
(iii) masses and mixings of quarks and leptons, as well as (iv)
neutrino oscillations, within a single predictive framework based
on SUSY grand unification.

In this regard, (a) the observed quantum numbers of the members of
a family, (b)  gauge coupling unification, (c) neutrino
oscillations, as well as (d) the likely need for leptogenesis
\cite{Yanagida} as a prelude to baryogenesis, together, suggest
that the SM gauge symmetry very likely emerges, near the GUT scale
$M_U \sim 2\times 10^{16}\ GeV$, from the spontaneous breaking of
a higher gauge symmetry that possesses the symmetry SU(4)-color
\cite{PS}. The higher gauge symmetry in 4D could be either SO(10)
\cite{SO(10)}, or an effective (presumably string derived)
symmetry G(224) = $SU(2)_L\times SU(2)_R\times SU(4)^c$ \cite{PS}.
The need for SU(4)-color arises because it provides: (a) the
right-handed neutrino ($\nu_R$) as a compelling member of each
family, (b) $B-L$ as a local symmetry, and (c) the Dirac mass of
$\nu_{\tau}$ in terms of the top-quark mass. These three
ingredients, together with the SUSY unification scale, seem to be
rather crucial \cite{JCPKeK} to an understanding of the neutrino
masses via the seesaw mechanism \cite{seesaw} and implementing
baryogenesis via leptogenesis \cite{Yanagida}. The theory thus
described may be viewed to have emerged in 4D from a string/M
theory near the string scale $M_{st} \gsim M_{GUT}$ $\footnote{The
relative advantages of SO(10) versus G(224) as effective 4D
symmetries involving the issues of coupling unification on the one
hand and doublet-triplet splitting on the other hand are discussed
in \cite{JCPKeK}.}$. It should of course possess weak scale
supersymmmetry so as to avoid unnatural fine tuning in the Higgs
mass and ensure gauge coupling unification.

A predictive framework based on the symmetry SO(10) or G(224), and
a minimal Higgs system which accomplish all these has been
proposed in Ref. \cite{BPW}, which we refer to as the BPW model.
This model describes the masses and mixings of all fermions
including neutrinos by making the simplifying assumption that the
fermion mass matrices are real and thus CP-conserving.
Notwithstanding this assumption, the framework is found to be
remarkably successful. In particular, it makes seven predictions
involving fermion masses, CKM elements and neutrino oscillations,
all in good accord with observations, to within 10\% (see
discussion in Sec. 2).

Now in general one would of course expect the entries in the
fermion mass matrices to have phases because the VEVs of the
relevant Higgs fields, and/or the effective Yukawa couplings, can
well be complex. These in turn can induce CP violation through the
SM/CKM interactions as well as through SUSY interactions involving
sfermion/gaugino loops.

The question arises: Can the BPW-framework of Ref. \cite{BPW},
based on the supersymmetric SO(10) or G(224)-symmetry, be
extended, by allowing for phases in the fermion mass matrices, so
as to yield net CP and flavor-violations, arising through both SM
and SUSY interactions, in accord with observations, while still
preserving its successes as regards fermion masses and neutrino
oscillations? This is the question we pursue in this paper.

As we will see, these four phenomena - (i) fermion masses, (ii)
neutrino oscillations, (iii) CP non-conservation, and (iv) flavor
violations - get intimately linked to each other within the SUSY
SO(10)/G(224) framework. Satisfying simultaneously the observed
features of all four phenomena within such a predictive framework
turns out, however, to be a non-trivial challenge to meet. One
main purpose of the present paper is to show that the answer to
the question raised above is in the affirmative. We defer the
discussion of our results and predictions in this regard to
section 5.

Since CP violation would have its origin, in our model, entirely
in the fermion mass matrices, and since the BPW framework
presented in Ref. \cite{BPW} has proven to be so successful in
describing fermion masses and neutrino oscillations, we first
briefly recall the salient features of this prior work in the next
section. In the following section we extend the same so as to
include CP violation.

\section{Fermion Masses and Neutrino Oscillations in G(224)/SO(10): A Brief
 Review of Prior Work}

The $3\times 3$ Dirac mass matrices for the four sectors
$(u,d,l,\nu)$ proposed in Ref. \cite{BPW} were motivated in part
by the notion that flavor symmetries \cite{Hall} are responsible
for the hierarchy among the elements of these matrices (i.e., for
``33"$\gg$``23"$\gg$``22"$\gg$``12"$\gg$``11", etc.), and in part
by the group theory of SO(10)/G(224), relevant to a minimal Higgs
system. Up to minor variants \cite{FN4}, they are as follows:
\begin{eqnarray}
\label{eq:mat}
\begin{array}{cc}
M_u=\left[
\begin{array}{ccc}
0&\epsilon'&0\\-\epsilon'&\zeta_{22}^u&\sigma+\epsilon\\0&\sigma-\epsilon&1
\end{array}\right]{\cal M}_u^0;&
M_d=\left[
\begin{array}{ccc}
0&\eta'+\epsilon'&0\\
\eta'-\epsilon'&\zeta_{22}^d&\eta+\epsilon\\0& \eta-\epsilon&1
\end{array}\right]{\cal M}_d^0\\
&\\
M_\nu^D=\left[
\begin{array}{ccc}
0&-3\epsilon'&0\\3\epsilon'&\zeta_{22}^u&\sigma-3\epsilon\\
0&\sigma+3\epsilon&1\end{array}\right]{\cal M}_u^0;& M_l=\left[
\begin{array}{ccc}
0&\eta'-3\epsilon'&0\\
\eta'+3\epsilon'&\zeta_{22}^d&\eta-3\epsilon\\0& \eta+3\epsilon&1
\end{array}\right]{\cal M}_d^0\\
\end{array}
\end{eqnarray}
These matrices are defined in the gauge basis and are multiplied
by $\bar\Psi_L$ on left and $\Psi_R$ on right. For instance, the
row and column indices of $M_u$ are given by $(\bar u_L, \bar c_L,
\bar t_L)$ and $(u_R, c_R, t_R)$ respectively. Note the
group-theoretic up-down and quark-lepton correlations: the same
$\sigma$ occurs in $M_u$ and $M_\nu^D$, and the same $\eta$ occurs
in $M_d$ and $M_l$. It will become clear that the $\epsilon$ and
$\epsilon'$ entries are proportional to $B-L$ and are
antisymmetric in the family space (as shown above). Thus, the same
$\epsilon$ and $\epsilon'$ occur in both ($M_u$ and $M_d$) and
also in ($M_\nu^D$ and $M_l$), but $\epsilon\rightarrow
-3\epsilon$ and $\epsilon'\rightarrow -3\epsilon'$ as
$q\rightarrow l$. Such correlations result in enormous reduction
of parameters and thus in increased predictiveness. Such a pattern
for the mass-matrices can be obtained, using a minimal Higgs
system ${\bf 45}_H,{\bf 16}_H,{\bf \overline {16}}_H \mbox{ and
}{\bf 10}_H $
 and a singlet $S$ of SO(10), through effective couplings as follows \cite{FN26} (see Ref.\cite{BPW} and \cite{JCPKeK} for details):
\begin{eqnarray}
\label{eq:Yuk} {\cal L}_{\rm Yuk} &=& h_{33}{\bf 16}_3{\bf
16}_3{\bf 10}_H  + [ h_{23}{\bf 16}_2{\bf 16}_3{\bf
10}_H(S/M) \nonumber \\
&+& a_{23}{\bf 16}_2{\bf 16}_3{\bf 10}_H ({\bf
45}_H/M')(S/M)^p+g_{23}{\bf 16}_2{\bf 16}_3{\bf 16}_H^d ({\bf
16}_H/M'')(S/M)^q] \nonumber \\ &+& \left[h_{22}{\bf 16}_2{\bf
16}_2{\bf 10}_H(S/M)^2+g_{22}{\bf 16}_2{\bf 16}_2 {\bf
16}_H^d({\bf 16}_H/M'')(S/M)^{q+1} \right] \nonumber \\ &+&
\left[g_{12}{\bf 16}_1{\bf 16}_2 {\bf 16}_H^d({\bf
16}_H/M'')(S/M)^{q+2}+ a_{12}{\bf 16}_1{\bf 16}_2 {\bf 10}_H({\bf
45}_H/M')(S/M)^{p+2} \right]~.
\end{eqnarray}
Typically we expect $M'$, $M''$ and $M$ to be of order $M_{\rm
string}$ or (possibly) of order $M_{GUT}$\cite{FN6}. The VEV's of
$\langle{\bf 45}_H\rangle$ (along $B-L$), $\langle{\bf
16}_H\rangle=\langle{\bf\overline {16}}_H\rangle$ (along standard
model singlet sneutrino-like component) and of the SO(10)-singlet
$\langle S \rangle$ are of the GUT-scale, while those of ${\bf
10}_H$ and of the down type SU(2)$_L$-doublet component in ${\bf
16}_H$ (denoted by ${\bf 16}_H^d$) are of the electroweak scale
\cite{BPW,FN7}. Depending upon whether $M'(M'')\sim M_{\rm GUT}$
or $M_{\rm string}$ (see \cite{FN6}), the exponent $p(q)$ is
either one or zero \cite{FN8}.

The entries 1 and $\sigma$ arise respectively from $h_{33}$ and
$h_{23}$ couplings, while $\hat\eta\equiv\eta-\sigma$ and $\eta'$
arise respectively from $g_{23}$ and $g_{12}$-couplings. The
$(B-L)$-dependent antisymmetric entries $\epsilon$ and $\epsilon'$
arise respectively from the $a_{23}$ and $a_{12}$ couplings.
[Effectively, with $\langle{\bf 45}_H\rangle\propto$ $B-L$, the
product ${\bf 10}_H\times{\bf 45}_H$ contributes as a {\bf 120},
whose coupling is family-antisymmetric.] The small entry
$\zeta_{22}^u$ arises from the $h_{22}$-coupling, while
$\zeta_{22}^d$ arises from the joint contributions of $h_{22}$ and
$g_{22}$-couplings. As discussed in \cite{BPW}, using some of the
observed masses as inputs, one obtains
$|\hat\eta|\sim|\sigma|\sim|\epsilon|\sim {\cal O}(1/10)$,
$|\eta'|\approx {\rm few} \times 10^{-3}$ and $|\epsilon'|\sim
2\times 10^{-4}$. The success of the framework presented in Ref.
\cite{BPW} (which sets $\zeta_{22}^u=\zeta_{22}^d=0$) in
describing fermion masses and mixings remains essentially
unaltered if $|(\zeta_{22}^u,\zeta_{22}^d)|\leq (1/3)(10^{-2})$
(say).

Such a hierarchical form of the mass-matrices, with $h_{33}$-term
being dominant, is attributed in part to flavor gauge
symmetry(ies) that distinguishes between the three families, and
in part to higher dimensional operators involving for example
$\langle{\bf 45}_H\rangle/M'$ or $\langle{\bf 16}_H\rangle/M''$,
which are suppressed by $M_{\rm GUT}/M_{\rm string}\sim 1/10$, if
$M'$ and/or $M''\sim M_{\rm string}$. As an example, introduce
just one $U(1)-$flavor gauge symmetry, together with a discrete
$Z_2-$symmetry $D$, with one singlet $S$. The hierarchical form of
the Yukawa couplings, in accord with Eqs. (\ref{eq:mat}) and
(\ref{eq:Yuk}), would follow, for the case of $p=1$, $q=0$, if,
for example, the $U(1)$ flavor charges are assigned as follows:$
\footnote{An alternative assignment with $p=1$, $q=1$ would arise
by choosing the charges: ($0,1,2,0,0,0,0$ and $-1$) for $({\bf
16}_3, {\bf 16}_2, {\bf 16}_1, {\bf 10}_H, {\bf 16}_H,
\overline{{\bf 16}}_H, {\bf 45}_H\ {\rm and}\ S)$ respectively.
Another variant would be to assign a flavor charge of 3 (or a+3 in
case of Eq. (\ref{eqn:charges})) to the first family ${\bf 16}_1$,
leaving all other charges as above. This would suppress ``13'',
``31'', ``12'', ``21'', and ``11'' entries by an order of
magnitude relative to the other cases.}$

\begin{equation}
\begin{array}{cccccccc}
\mathbf{16}_3  & \mathbf{16}_2 & \mathbf{16}_1 & \mathbf{10}_H &
\mathbf{16}_H & \overline{\mathbf{16}}_{H} &
\mathbf{45}_H & \mathbf{S} \\
a & a+1 & a+2 & -2a & -a-1/2 & -a & 0 & -1
\end{array}.
\label{eqn:charges}
\end{equation}
All the fields are assumed to be even under the discrete symmetry
$D$, except for $\mathbf{16}_{H}$ and $\overline{\mathbf{16}}_{H}$
which are odd. It is assumed that other fields are present that
would make the $U(1)$ symmetry anomaly-free. With this assignment
of charges, one would expect $|\zeta_{22}^{u,d}|\sim (\langle
S\rangle/M)^2$; one may thus take $|\zeta_{22}^{u,d}|\sim
(1/3)\times 10^{-2}$ without upsetting the success of
Ref.~\cite{BPW}. In the same spirit, one would expect
$|\zeta_{13}, \zeta_{31}|\sim (\langle S\rangle/M)^2\sim 10^{-2}$
and $|\zeta_{11}|\sim (\langle S\rangle/M)^4\sim 10^{-4}$ (say).
where $\zeta_{11}$, $\zeta_{13}$, and $\zeta_{31}$ denote the
``11'', ``13'', and ``31'', elements respectively. The value of
$a$ can get fixed by the presence of other operators (see Ref.
\cite{JCPErice}).

The Majorana masses of the RH neutrinos arise from effective
couplings of the form ${\cal L}_{\rm Maj}= f_{ij}{\bf 16}_i{\bf
16}_j\overline{\bf 16}_H \overline{\bf 16}_H/M$, where the
$f_{ij}$'s include appropriate powers of $\langle S/M\rangle$ in
accord with flavor-charge assignments of ${\bf 16}_i$ and ${\bf
\overline{16}_H}$, and $M$ is expected to be of order string or
reduced Planck-scale. From the flavor-charge assignments given in
Eq. (\ref{eqn:charges}), it is clear that the ``33'' entry is
leading and the other entries in the Majorana mass matrix have a
relative hierarchy which is $\it{identical}$ to that in the Dirac
mass matrices. We refer the reader to \cite{BPW} and \cite{JCPKeK}
for a more detailed discussion of the neutrino sector. Here, we
list only the results.

Ignoring possible phases in the parameters and thus the source of
CP violation for a moment, and also setting $\zeta_{22}^d =
\zeta_{22}^u = 0$, as was done in Ref. \cite{BPW}, the parameters
$(\sigma,\eta, \epsilon, \epsilon',\eta', {\cal M}_u^0,\ {\rm
and}\ {\cal M}_d^0)$ can be determined by using, for example,
$m_t^{\rm phys}=174$ GeV, $m_c(m_c)=1.37$ GeV, $m_s(1\mbox{
GeV})=110-116$ MeV, $m_u(1\mbox{ GeV})=6$ MeV, and the observed
masses of $e$, $\mu$, and $\tau$ as inputs. One is thus led, {\it
for this CP conserving case}, to the following fit for the
parameters, and the associated predictions \cite{BPW}:
\begin{eqnarray}
\label{eq:fit}
\begin{array}{l}
\ \sigma\approx 0.110, \quad \eta\approx 0.151, \quad
\epsilon\approx -0.095,
 \quad |\eta'|\approx 4.4 \times 10^{-3},\\
\begin{array}{l}
\epsilon'\approx 2\times 10^{-4},\quad {\cal M}_u^0\approx
m_t(M_X)\approx 100 \mbox{ GeV},\quad {\cal M}^0_d\approx
m_{\tau}(M_X)\approx 1.1 \mbox{ GeV}.
\end{array}
\end{array}
\end{eqnarray}
These output parameters remain stable to within 10\% corresponding
to small variations ($\lsim 10$\%) in the input parameters of
$m_{t}$, $m_{c}$, $m_{s}$, and $m_{u}$. These in turn lead to the
following predictions for the quarks and light neutrinos
\cite{BPW}, \cite{JCPKeK}:
\begin{eqnarray}
\label{eq:pred}
\begin{array}{l}
\ \ m_b(m_b)\approx4.9\mbox{ GeV},\quad\sqrt{\Delta m_{23}^2}
\approx
m(\nu_3)\approx\mbox{(1/24 eV)(1/2-2)},\quad V_{cb}\approx 0.044, \\
\begin{array}{l}
\ \sin^2 2\theta^{\rm osc}_{\nu_{\mu}\nu_{\tau}}\approx
0.98-0.995,
\mbox{   (for $m(\nu_2)/m(\nu_3)\approx 1/10-1/5$),}\\
\begin{array}{l}
|V_{us}|\approx 0.20,\quad \left|\frac{V_{ub}}{V_{cb}} \right|\approx 0.07,\quad m_d(\mbox{1 GeV})\approx \mbox{8 MeV}\\
\end{array}
\end{array}
\end{array}
\end{eqnarray}

It has been noted \cite{JCPErice,JCPKeK} that small non-seesaw
contribution to $\nu_L^e\nu_L^{\mu}$ mass term ($\sim $ few
$\times 10^{-3}$ eV) which can arise through higher dimensional
operators, but which have been ignored in the analysis given
above, can lead quite plausibly to large $\nu_e-\nu_{\mu}$
oscillation angle in accord with the LMA MSW solution for the
solar neutrino problem. Leaving aside therefore the question of
the $\nu_e-\nu_{\mu}$ oscillation angle, it seems quite remarkable
that all seven predictions in Eq. (\ref{eq:pred}) agree with
observations to within 10\%. Particularly intriguing is the
$(B-L$)-dependent group-theoretic correlation between $V_{cb}$ and
$\theta_{\nu_{\mu}\nu_{\tau}}^{osc}$, which explains
simultaneously why one is small ($V_{cb}$) and the other is so
large ($\theta_{\nu_{\mu}\nu_{\tau}}^{osc}$) \cite{BPW,JCPKeK}.
That in turn provides some confidence in the ${\it gross\
pattern}$ of the Dirac mass matrices presented above and motivates
the study of CP and flavor violations within the same framework.
This is what we do in the next section.

\section{Phases in the Fermion Mass
Matrices: The Origin of CP violation }

In the work of Ref. \cite{BPW} reviewed above, the parameters
($\sigma,\ \epsilon,\ \eta,\ \epsilon',\ \eta'$ etc.) entering
into the fermion mass matrices were assumed to be real, for
simplicity, and thereby (at least) the SM interactions were
rendered CP-conserving$\footnote{modulo the contribution from the
strong CP parameter $\Theta$}$. Noting that the VEVs of the Higgs
fields$\footnote{For instance, consider the superpotential for
${\bf 45}_H$ only: W(${\bf 45}_H$)= $M_{{\bf 45}}({\bf
45}_H)^2+``\lambda ({\bf 45}_H)^4$''/$M$, which yields (setting
$F_{{\bf 45}_H}$=0), either $\langle {\bf 45}_H\rangle$ = 0, or
$\langle {\bf 45}_H\rangle^2=-(2M_{{\bf 45}} M/``\lambda$'').
Assuming that ``other physics'' would favor $\langle {\bf
45}_H\rangle \ne 0$, we see that $\langle {\bf 45}_H\rangle$ would
be pure imaginary, if the quantity in the brackets is positive
with all parameters being real. In a coupled system, it is
conceivable that $\langle {\bf 45}_H\rangle$ in turn would induce
phases (other than 0 and $\pi$) in some of the other VEVs as well,
and may itself become complex rather than pure imaginary. }$
and/or the effective Yukawa couplings can well be complex,
however, we now propose to extend the SO(10)/G(224) framework
reviewed above to include CP violation by allowing for these
parameters to have phases.

Given the empirical constraints on (i) CP and flavor violations,
as well as (ii) fermion masses and (iii) neutrino oscillations, on
the one hand, and (iv) the group-theoretical constraints of the
SO(10)/G(224) framework on the other, it is of course not at all
clear, a priori, whether any choice of phases and variations in
the parameters of the fermion mass matrices presented above can
yield {\it observed} CP and flavor-violations, and simultaneously
preserve the successes of the framework of \cite{BPW} as regards
fermion masses and neutrino oscillations. That is the issue we now
explore. We choose to diagonalize the quark mass matrices $M_u$
and $M_d$ at the GUT scale $\sim 2\times 10^{16}$ GeV, by
bi-unitary transformations - i.e.
 \begin{eqnarray}
 \label{eq:xdxu}
M_d^{diag}\ =\ X_L^{d\dagger}M_d X_R^{d}\ {\rm and} \ M_u^{diag}\
=\ X_L^{u\dagger}M_u X_R^{u}
\end{eqnarray}
with phases of $q_{L,R}^i$ chosen such that the eigenvalues are
real and positive and that the CKM matrix $V_{CKM}$ (defined
below) has the Wolfenstein form \cite{Wolfenstein}). Utilizing
the hierarchical nature of the mass matrices, one can obtain
(approximate) analytic expressions for the diagonalizing
matrices. They are:
\begin{eqnarray}
\label{eq:xdl}
\begin{array}{cc}
X_L^d \simeq \left[
\begin{array}{ccc}
e^{-i(\phi_{\eta-\epsilon} )}& |\eta'/X_d|e^{-i(\phi_{\eta-\epsilon} +\zeta_{us})}&\eta'|\eta-\epsilon|e^{-i(\phi_{\eta-\epsilon} -\zeta_{33}^d)}\\
-|\eta'/X_d|e^{i(\phi_{\eta+\epsilon} +\phi_{X_d})}& e^{i(\phi_{\eta+\epsilon} +\phi_{X_d}-\zeta_{us})}& |\eta+\epsilon|e^{i(\phi_{\eta+\epsilon} +\zeta_{33}^d)}\\
|\eta'/X_d||\eta+\epsilon|e^{i(\phi_{X_d})}-Y &
-|\eta+\epsilon|e^{i(\phi_{X_d}-\zeta_{us})}& e^{i\zeta_{33}^d}
\end{array}\right]
\end{array}
\end{eqnarray}
\begin{eqnarray}
\label{eq:xdr}
\begin{array}{cc}
X_R^d \simeq \left[
\begin{array}{ccc}
e^{i(\phi_{\eta+\epsilon} +\phi_{X_d})}&|\eta'/X_d|e^{i(\phi_{\eta+\epsilon} +\phi_{X_d}-\zeta_{us})}&\eta'|\eta+\epsilon|e^{i(\phi_{\eta+\epsilon} +\zeta_{33}^d)}\\
-|\eta'/X_d|e^{-i(\phi_{\eta-\epsilon} )}& e^{-i(\phi_{\eta-\epsilon} +\zeta_{us})}& |\eta-\epsilon|e^{-i(\phi_{\eta-\epsilon} -\zeta_{33}^d)}\\
|\eta'/X_d||\eta-\epsilon|& -|\eta-\epsilon|e^{-i\zeta_{us}}&
e^{i\zeta_{33}^d}
\end{array}\right]
\end{array}
\end{eqnarray}

Here $\phi_{\eta\pm\epsilon}\equiv arg(\eta\pm\epsilon)$, that is,
($\eta\pm\epsilon\equiv|\eta\pm\epsilon|\
e^{i\phi_{\eta\pm\epsilon}}$); $Y \equiv
\eta'|\eta-\epsilon|e^{-i\zeta_{ud}}$ and\\
 $X_d \equiv \ -|\eta^2-\epsilon^2|
+|\zeta_{22}^d|e^{-i(\phi_{\eta+\epsilon}+\phi_{\eta-\epsilon}-\phi_{\zeta_{2d}})}\equiv
|X_d|e^{i\phi_{X_d}}$, where
$\zeta_{22}^d\equiv|\zeta_{22}^d|e^{i\phi_{\zeta_{2d}}}$. The
corresponding matrices $X_{L,R}^u$ for diagonalizing the up sector
are obtained from above with the substitutions :
$\eta\rightarrow\sigma$; $\zeta_{22}^d\rightarrow\zeta_{22}^u $;
$(\eta'\pm\epsilon')\rightarrow\pm\epsilon'$. Thus
$\phi_{\eta\pm\epsilon}$ are replaced by
$\phi_{\sigma\pm\epsilon}\equiv arg(\sigma\pm\epsilon)$; and $X_d$
by $X_u\equiv\ -|\sigma^2 -\epsilon^2|
+|\zeta_{22}^u|e^{-i(\phi_{\sigma+\epsilon}+\phi_{\sigma-\epsilon}-\phi_{\zeta_{2u}})}\equiv
|X_u|e^{i\phi_{X_u}}$. Given the definitions of $\phi_{X_d}$ and
$\phi_{X_u}$ as above, we have
$$\zeta_{33}^d\simeq
(\phi_{X_d}-\phi_{\eta-\sigma}+\phi_{\eta+\epsilon})+R \sin
\Omega;\quad
\gamma\equiv(\phi_{\eta+\epsilon}+\phi_{\eta-\epsilon})-(\phi_{\sigma+\epsilon}+
\phi_{\sigma-\epsilon})+\phi_{\epsilon'},$$ where
\begin{eqnarray}
R &\equiv&
|\epsilon'/X_u|/|\eta'/X_d|\approx\sqrt{m_u/m_c}\sqrt{m_s/m_d}\approx\
0.3; \nonumber \\
\beta_\Omega &\approx& \ R(\sin\Omega/\Omega),~
\Omega\equiv(\phi_{X_d}-\phi_{X_u})+\gamma;\nonumber \\
\zeta_{cb}&\simeq&
arg[e^{i(\gamma-\phi_{X_u})}\{|\eta+\epsilon|-\mid\sigma+\epsilon|
e^{i(\phi_{\sigma+\epsilon}-\phi_{\eta+\epsilon})}\}]; \nonumber
\\ \zeta_{us}&\approx& -R\ \sin\Omega[1-R\
\cos\Omega]^{-1}.\nonumber
\end{eqnarray}

As mentioned above, using observed fermion masses and mixings
\cite{BPW}, we obtain: $ |\epsilon'|\sim\ 1/10\ |\eta'|$, with
$|\eta'|\sim$ (few)$\times10^{-3}\ll
(|\eta|\sim|\epsilon|\sim|\sigma|\sim1/10)\ll1$. In writing Eqns.
(\ref{eq:xdl}) and (\ref{eq:xdr}), we have not displayed for
simplicity of writing, small correction terms $({\cal O}
(\epsilon^{2},\eta^{2}))$, which are needed to preserve unitarity.
We have also not displayed small phases of order
$|\eta'\epsilon'/X_u X_d|\times$ sin$\ \Omega\sim 1/100$,$\
|\epsilon'/\eta'|\sim 1/10$ and  $R \sin\Omega\sim 1/10.$ Our
results to be presented, that are based on exact numerical
calculations,
 however incorporate these small corrections.

The CKM elements in the Wolfenstein basis are given by the matrix
$V_{CKM}=e^{-i\alpha}(X_L^{u \dagger} X_L^d)$, where
$\alpha=(\phi_{\sigma-\epsilon}-\phi_{\eta-\epsilon})-(\phi_{\epsilon'}-\phi_{\eta'+\epsilon'})$,
where without loss of generality (given $|\eta'|\gg|\epsilon'|$),
we can choose $\phi_{\eta'+\epsilon'}\approx0$. To a good
approximation, the CKM elements are given by:
\begin{eqnarray}
\label{eq:ckmelem}
\begin{array}{l}
\ \ \ \ V_{ud}\approx V_{cs}\approx V_{tb}\approx 1\\
\begin{array}{l}
\ \ \ V_{us}\approx
\mid|\eta'/X_d|-|\epsilon'/X_u|e^{i\Omega}\mid\
\approx\  -V_{cd} \\
\begin{array}{l}
\ \ V_{cb}\approx \mid
e^{i(\gamma-\phi_{X_u})}\{\mid\eta+\epsilon\mid-\mid\sigma+\epsilon\mid
e^{i(\phi_{\sigma+\epsilon}-\phi_{\eta+\epsilon})}\}\mid\ \approx\ -V_{ts} \\
\begin{array}{l}
\ V_{ub}\approx [\eta'\mid\eta-\epsilon\mid-\mid\epsilon'/X_u\mid
e^{i(\gamma-\phi_{X_u})}\{|\eta+\epsilon|-|\sigma+\epsilon|e^{-i(\phi_{\sigma+\epsilon}-\phi_{\eta+\epsilon})}\}]\times
e^{i[\Omega(1+\beta_{\Omega})-\zeta_{cb}]} \\
\begin{array}{l}
V_{td}\approx[\mid\eta'/X_d\mid
e^{i(\phi_{X_d})}\{\mid\epsilon+\eta\mid-\mid\sigma+\epsilon\mid
e^{-i(\phi_{\sigma+\epsilon}-\phi_{\eta+\epsilon})}\}-\eta'|\eta-\epsilon|]\times
e^{-i[\Omega(1+\beta_{\Omega})-\zeta_{cb}]}
\end{array}
\end{array}
\end{array}
\end{array}
\end{array}
\end{eqnarray}

\noindent Note that the CKM elements have the desired Wolfenstein
form with only $V_{ub}$ and $V_{td}$ being complex and the others
being real to a good approximation. $\zeta_{cb}$ defined above is
just the argument of the expression within the bars for $V_{cb}$.
One can check that to a good approximation, (neglecting the
$\eta'|\eta-\epsilon|$ term for $V_{td}$ that causes $<$ 10\%
error), the phase of $V_{td}$ is given by \linebreak
$\phi_{td}\equiv Arg (V_{td}) \approx -R \sin \Omega$, and
$|V_{td}|\approx |\eta'/X_d||V_{cb}^{*}|\approx\ \sqrt{m_d/m_s}\
|V_{cb}|$, and similarly $|V_{ub}|\approx\ \sqrt{m_u/m_c}\
|V_{cb}|$.

Before presenting the results of a certain fit and the
corresponding predictions, we need to first discuss SUSY CP and
flavor violations in the presence of phases in the fermion mass
matrices. This is done in the next section.

\section{SUSY CP and Flavor Violations}
Our procedure for dealing with SUSY CP and flavor violations may
be summarized by the following set of considerations:

1) As is well known, since the model is supersymmetric,
non-standard CP and flavor violations would generically arise in
the model through sfermion/gaugino quantum loops involving scalar
$(mass)^2$ transitions. The latter can either preserve chirality
(as in $\tilde{q}^i_{L,R}\rightarrow \tilde{q}^j_{L,R}$) or flip
chirality (as in $\tilde{q}^i_{L,R}\rightarrow
\tilde{q}^j_{R,L}$). Subject to our assumption on SUSY breaking
(specified below), it would turn out that these scalar $(mass)^2$
parameters get completely determined within our model by the
fermion mass-matrices, and the few parameters of SUSY breaking.

2) $\bf{SUSY\  Breaking:}$ We assume that SUSY breaking is
communicated to the SM sector by messenger fields which have large
masses of order $M^{*}$, where $M_{GUT}\lsim M^{*}\approx
M_{string}$, such that the soft parameters are flavor-blind, and
family-universal at the scale $M^{*}$. A number of well motivated
models of SUSY breaking, e.g., those based on mSUGRA
\cite{msugra}, gaugino-mediation \cite{gauginomed}, anomalous
$U(1) - D$ term \cite{anomU(1), faraggiJCP}, combined with
dilaton-mediation \cite{faraggiJCP,dilaton}, or possibly a
combination of some of these mechanisms, do in fact induce such a
breaking. While for the first two cases \cite{msugra, gauginomed}
we would expect extreme squark degeneracy (ESD) i.e. $\kappa\equiv
|m^2(\tilde{q_i})- m^2(\tilde{q}_j)|/m^2(\tilde{q})_{AV}\ \ll\
10^{-3}$ (say) at the scale $M^{*}$, for the third case
\cite{anomU(1),faraggiJCP}, one would expect intermediate squark
degeneracy (ISD) i.e. $\kappa\sim 10^{-2}(1-1/3)$ at $ M^{*}$. For
the sake of generality, we would initially allow both
possibilities, $\kappa$ = 0 (ESD), and $\kappa\sim10^{-2}(1-1/3)$
(ISD) at $M^{*}$.

In an extreme version of universality, analogous to  CMSSM, the
SUSY sector of the model would introduce only five parameters at
the scale $M^{*}$:
\begin{center}
$m_o, m_{1/2}, A_o, \tan\beta\ {\rm and}\ sgn(\mu).$
\end{center}
In some cases, $A_o$ can be zero or extremely small ($\lsim$ 1GeV)
at $M^{*}$ as in \cite{gauginomed} and \cite{faraggiJCP}. For most
purposes we will adopt this restricted version of SUSY breaking,
including the vanishing of $A_o$ at $M^{*}$. However, our results
will be essentially unaffected even if $A_o$ is non-zero ($\sim$
500 GeV, say) but real (see remarks later). We will not insist on,
but will allow for, Higgs-squark-slepton universality, which does
not hold, for example, in the string-derived model of
\cite{faraggiJCP}. In spite of flavor-preservation at a high scale
$M^{*}$, SUSY-induced flavor-violation would still arise at the
electroweak scale through renormalization group running of the
sfermion masses and the $A$-parameters from $M^{*}\rightarrow
M_{GUT}\rightarrow m_W$, as specified below. Although the premises
of our model as regards the choice of universal SUSY parameters
coincide with that of CMSSM, as we will see, owing to the presence
of GUT-scale physics in the interval $M^{*}\rightarrow M_{GUT}$,
SUSY CP and flavor violations in our model (evaluated at the
electroweak scale) would be significantly enhanced compared to
that in CMSSM (or even CMSSM with right-handed neutrinos). This
difference provides some distinguishing features of our model.

3) $\bf{Flavor\ Violation\ due\ to\ RG\ Running\ of\ Scalar\
Masses\ from\ M^{*}\ to\ M_{GUT}}$

For MSSM embedded into SO(10) above the GUT scale, there
necessarily exist heavy color-triplet Higgs fields which couple to
fermions through the coupling $h_{t}{\bf 16}_3{\bf 16}_3{\bf 10}_H
$, while there exist heavy doublets for both SO(10) and G(224)
which also couple to fermions owing to the mixing of ${\bf 10}_H$
with ${\bf 16}_H$ (see \cite{BPW}). (Here $h_t$ stands for
$h_{33}$ of Eq. (\ref{eq:Yuk})). These couple to $\tilde{b}_L $
and $\tilde{b}_R$ with the large top quark Yukawa coupling $h_t$.
The heavy triplets and doublets possess masses of order $M_{GUT}$.
One can verify (see \cite{BarbieriHallStrumia}) that the evolution
of RG equations for squark masses involving such couplings
suppress $\tilde{b}_L$ and $\tilde{b}_R$ masses significantly
compared to those of $\tilde{d}_{L,R}$ and $\tilde{s}_{L,R}$. Note
that left--right symmetry implies equal shifts in $\tilde{b}_L$
and $\tilde{b}_R$ masses arising from GUT scale physics in the
momentum range $M_{GUT} \leq \mu \leq M^*$. Such differential mass
shifts i.e.- ${(\hat{m}_{3}^2-\hat{m}_{1,2}^2)}_{L,R}\equiv
\Delta\hat{m}_{\tilde{b}_{L,R}}^2$, for the embedding of MSSM into
SO(10), are found to be (with $A_o = 0$):
\begin{eqnarray}
\label{eq:deltambr} \Delta\hat{m}_{\tilde{b}_{R}}^2 =
\Delta\hat{m}_{\tilde{b}_{L}}^2 \approx
-\bigl(\frac{30m_o^2}{16\pi^2}\bigr) h_t^2\ ln(M^{*}/M_{GUT})
\equiv -(m_o^2/4)\xi~.
\end{eqnarray}
The hat signifies GUT scale values. Here $m_o$ denotes the
(approximately) degenerate mass of squarks at the scale $M^{*}$.
We have set $h_t^2 =1/2$; we expect $M^{*}/M_{GUT}\ \sim$ (3 to
10), say, and thus, $\xi\equiv ln(M^{*}/M_{GUT})/2.6 \approx (0.4\
to\ 0.9)$. For the case of MSSM embedded into G(224), which
provides the heavy doublet, but not the triplets, the factor 30 in
Eq. (\ref{eq:deltambr}) should be replaced by 12.

Having diagonalized the quark mass-matrices $M_d^{(0)}$ and
$M_u^{(0)}$ at the GUT scale by matrices as in Eq.
(\ref{eq:xdxu}), SUSY flavor violation may be assessed by imposing
the parallel transformations on the squark $(mass)^2$ matrices
$((\tilde{M}^{(0)}_d)_{LL/RR})$ defined in the gauge basis, i.e.,
by evaluating $X_L^{d\dagger}(\tilde{M}^{(0)}_d)_{LL}X_L^d$ and
$X_R^{d\dagger}(\tilde{M}^{(0)}_d)_{RR}X_R^d$, and similarly for
the up sector. Following discussion on SUSY breaking, the
off--diagonal elements (in the gauge basis) and all chirality
flipping elements are set to be zero - i.e
$(\tilde{M}^{(0)}_{ij})_{LL/RR}= 0\ (i\neq j)$ and
$(A^0_{ij})_{LR} = 0$ - at the scale $M^{*}$. Once squarks are non
degenerate at $M_{GUT}$ owing to the mass-shift of
$\tilde{b}_{L,R}$ as in Eq. (\ref{eq:deltambr}), the
transformations mentioned above induce off-diagonal elements with
squarks being in the SUSY basis. For the down squark mass matrices
(evaluated at the GUT scale), these off diagonal elements are
found to be:

\begin{eqnarray}
\label{eq:deltaij}
\begin{array}{l}
\ \ \ \ \ \hat\delta_{LL}^{12}(M_{GUT})\simeq
[\kappa_{ISD}^{12}+(\Delta\hat{m}_{\tilde{b}_{L}}^2/m_{sq}^2)(-|\eta'/X_d||\epsilon+\eta|^2+\eta'|\epsilon^2-\eta^2|e^{i\phi_{X_d}})\}]e^{-i\phi_{td}}\\
\quad\quad\quad\quad\quad\quad\  \ \approx [\kappa_{ISD}^{12} + 1.5\times 10^{-4}\xi](m_o^2/m_{sq}^2)e^{-i\phi_{td}}\\
\\
\begin{array}{l}
\ \ \ \ \hat\delta_{RR}^{12}(M_{GUT})\simeq
[\kappa_{ISD}^{12}+(\Delta\hat{m}_{\tilde{b}_{R}}^2/m_{sq}^2)(-|\eta'/X_d||\epsilon-\eta|^2+\eta'|\epsilon^2-\eta^2|e^{-i\phi_{X_d}})\}]e^{-i\phi_{td}}\\
\quad\quad\quad\quad\quad\quad\ \ \approx[\kappa_{ISD}^{12}+
3\xi\times
10^{-3}-10^{-5}(\xi)e^{-i\phi_{X_d}}](m_o^2/m_{sq}^2)e^{-i\phi_{td}}\\
\\
\begin{array}{l}
\ \ \ \hat\delta_{LL}^{13}(M_{GUT})\ \simeq
(\Delta\hat{m}_{\tilde{b}_{L}}^2/m_{sq}^2)[-\eta'|\eta-\epsilon|e^{i\zeta_{33}^d}+|\eta'/X_d||\eta+\epsilon|e^{i(\zeta_{33}^d-\phi_{X_d})}]\\
\quad\quad\quad\quad\quad\quad\ \approx[(2.5\xi)\times
10^{-4}e^{i\zeta_{33}^d}-(2.5\xi)\times
10^{-3}e^{i(\zeta_{33}^d-\phi_{X_d})}](m_o^2/m_{sq}^2)\\
\\
\begin{array}{l}
\ \ \hat\delta_{RR}^{13}(M_{GUT})\ \simeq
(\Delta\hat{m}_{\tilde{b}_{R}}^2/m_{sq}^2)[-\eta'|\eta+\epsilon|e^{i(\zeta_{33}^d-\phi_{X_d})}+|\eta'/X_d||\eta-\epsilon|e^{i\zeta_{33}^d}]\\
\quad\quad\quad\quad\quad\quad\ \approx -[(1.25\xi)\times
10^{-2}e^{i\zeta_{33}^d}](m_o^2/m_{sq}^2)\\
\\
\begin{array}{l}
\ \hat\delta_{LL}^{23}(M_{GUT})\simeq
(\Delta\hat{m}_{\tilde{b}_{L}}^2/m_{sq}^2)[-|\eta+\epsilon|e^{i(\zeta_{33}^d-\phi_{X_d}+\phi_{td})}]\\
\quad\quad\quad\quad\quad\ \ \approx[(1.25\xi)\times
10^{-2}e^{i(\zeta_{33}^d-\phi_{X_d}+\phi_{td})}](m_o^2/m_{sq}^2)\\
\\
\begin{array}{l}
\hat\delta_{RR}^{23}(M_{GUT})\simeq(\Delta\hat{m}_{\tilde{b}_{R}}^2/m_{sq}^2)[-|\eta-\epsilon|e^{i(\zeta_{33}^d+\phi_{td})}]\\
\quad\quad\quad\quad\quad\ \approx[(6.2\xi)\times
10^{-2}e^{i(\zeta_{33}^d+\phi_{td})}](m_o^2/m_{sq}^2)\\
\end{array}
\end{array}
\end{array}
\end{array}
\end{array}
\end{array}
\end{eqnarray}

\noindent Here $\kappa_{ISD}^{12}\equiv\
[(m_1^{(0)2}-m_2^{(0)2})]/m_{sq}^2(|\eta'/X_d|)\sim\pm(2\times
10^{-3})(1-1/3)$; this term would be present for the case of
intermediate squark degeneracy (ISD), corresponding to small
$(\sim 10^{-2}(1-1/3))$ squark non-degeneracy at the scale
$M^{*}$, as in models of Ref. \cite{anomU(1),faraggiJCP}. From now
on, for the sake of concreteness, we drop this
term,$\footnote{Note that the case of ISD ($\kappa_{ISD}^{12}\sim
(2\times 10^{-3})(1-1/3)\ne 0$) would make a difference only for
the case of $K^{o}-\overline{K^{o}}$ transitions - that is, for
$\Delta m_K$ and $\epsilon_K$.}$ setting $\kappa_{ISD}^{12}=0$. In
above$\ \phi_{td}\approx-\ |\epsilon'/X_u|/|\eta'/X_d|\
\sin\Omega\sim(-1/3)\ \sin\Omega\sim (-1/6)$(say). The hat on top
signifies GUT scale values, and
$\hat{\delta}_{LL/RR}^{ij}\equiv(\hat{\Delta}_{LL/RR}^{ij})/m_{sq}^2$,
where $\hat{\Delta}_{LL}^{ij}$ denotes the $(mass)^2$ parameter
for $\tilde{q}_{jL}\rightarrow\tilde{q}_{iL}$ transition in the
SUSY basis. Here, $m_{sq}$ denotes the average mass of the
$\tilde{d}_{L,R}$ and $\tilde{s}_{L,R}$ squarks, which remain
nearly degenerate( to 1\% or better) even at the weak scale. For
each $\hat{\delta}_{LL/RR}^{ij}$ we have exhibited approximate
numerical values by inserting values of the parameters $\eta,\
\sigma,\ \epsilon\ \eta'$ etc. for some typical fits (as in Eqs.
(\ref{eq:fit}) and (\ref{eq:fitA})) to indicate their typical
values.

Assuming for simplicity, universality of scalar masses $m_o$ (of
the first two families) and of the gaugino masses $m_{1/2}$ at the
GUT scale, the physical masses of squarks of the first two
families and of the gluino are given by:
\begin{eqnarray}
m_{sq}^2\approx m_o^2 + 7.2 ~m_{1/2}^2;\quad\ m_{\tilde{g}}\approx
2.98 ~m_{1/2}~.
\end{eqnarray}
This result is rather insensitive to the mass shifts of
$\tilde{b}_{L,R}$. Using the above relations we get $\rho_X\equiv
(m_o^2/m_{sq}^2)\simeq 1-0.8x\approx$ (0.84, 0.76, 0.5 and 0.2)
for $x\equiv m_{\tilde{g}}^2/m_{sq}^2$=(0.2, 0.3, 0.6 and 1),
which enters into all the $\hat{\delta}^{ij}$-elements in Eq.
(\ref{eq:deltaij}).

We remind the reader that the elements
$\hat{\delta}^{ij}_{LL,RR}$, induced solely through GUT scale
physics being relevant in the interval $M^{*}\rightarrow M_{GUT}$,
would be absent in a general CMSSM or MSSM, and so would the
associated CP and flavor violations.

4) $\bf{Flavor\ Violation\ Through\ RG\ Running\ From\ M_{GUT}\
to\ m_W\ in\ MSSM}:$ It is well known that, even with universal
masses at the GUT scale, RG running from $M_{GUT}$ to $m_W$ in
MSSM, involving contribution from the top Yukawa coupling, gives a
significant correction to the mass of $\tilde{b}'_L =
V_{td}\tilde{d}_L+V_{ts}\tilde{s}_L+V_{tb}\tilde{b}_L$, which is
not shared by the mass-shifts of $\tilde{b}_R, \tilde{d}_{L,R}$
and $\tilde{s}_{L,R}$. This in turn induces flavor violation.
Here, $ \tilde{d}_L, \tilde{s}_L$ and $\tilde{b}_L$ are the SUSY
partners of the physical $d_L, s_L$ and $b_L$ respectively. The
differential mass shift of $\tilde{b}'_L$ arising as above, may be
expressed by an effective Lagrangian \cite{carena}: $\Delta{\cal
L} = -(\Delta m_L^{'2})\tilde{b}_L^{'*}\tilde{b}'_L$,
where$\footnote{Note that strictly speaking Eq.
(\ref{eq:deltam'l}) holds if the soft parameters are universal at
the GUT-scale. However, the correction to this expression due to
RG running from $M^{*}$ to $m_W$ would be rather small, being a
correction to a correction.}$
\begin{eqnarray}
\label{eq:deltam'l} \Delta m_L^{'2}=-3/2 m_o^2\eta_t + 2.3 A_o
m_{1/2} \eta_t(1-\eta_t) - (A_o^2/2)\eta_t(1-\eta_t) +
m_{1/2}^2(3\eta_t^2-7\eta_t)~.
\end{eqnarray}

Here $\eta_t = (h_t/h_f)\approx (m_t/v \sin\beta)^2(1/1.21)\approx
0.836$ for tan$\beta$ = 3. Numerically, setting$\footnote{Although
we have put $A_o$ = 0 (for concreteness), note that $\Delta
m^{'2}_L$ would typically get only a small correction ($\lsim$
5\%), even if $A_o$ were non-zero ($\lsim$ 1 TeV), with
$m_o\approx (0.7-1)$ TeV and $m_{1/2}\approx (200-300)$ GeV,
say.}$ $A_o$ = 0, Eq. (\ref{eq:deltam'l}) yields: $ (\Delta
m_L^{'2}/m_{sq}^2)\approx -(0.40,\ 0.34,\ 0.26,\ 0.22)$ for $x =
m_{\tilde g}^2/m_{sq}^2\approx (0.1,\ 0.4,\ 0.8,\ 1.0)$.
Expressing $\tilde{b}'_L$ in terms of down-flavor squarks in the
SUSY basis as above, Eq. (\ref{eq:deltam'l}) yields new
contributions to off diagonal squark mixing. Normalizing to
$m_{sq}^2$, they are given by
\begin{eqnarray}
\label{eq:delta'll} \delta_{LL}^{'(12,13,23)} =\bigl(\frac{\Delta
m_L^{'2}}{m_{sq}^2}\bigr)(V_{td}^{*}V_{ts},\ V_{td}^{*}V_{tb},\
V_{ts}^{*}V_{tb})~.
\end{eqnarray}

The net squark $(mass)^2$ off-diagonal elements at $m_W$ are then
obtained by adding the respective GUT-scale contributions from
Eqs. (\ref{eq:deltaij}) to that from Eq. (\ref{eq:delta'll}). They
are:
\begin{eqnarray}
\label{eq:deltallrr}
\delta_{LL}^{ij}=\hat\delta_{LL}^{ij}+\delta_{LL}^{'ij};\quad\
 \delta_{RR}^{ij}=\hat\delta_{RR}^{ij}
\end{eqnarray}

From the expressions given above (Eqs. (\ref{eq:deltaij}) and
(\ref{eq:delta'll})), it follows that for a given choice of the
SUSY-parameters (i.e. $m_o,\ m_{1/2}$ or equivalently $m_{sq}$ and
$m_{\tilde g}$), SUSY CP and flavor violations are
$\it{completely\ determined}$ within our model by parameters of
the fermion mass-matrices. This is the reason why within a
quark-lepton unified theory as ours, SUSY CP and flavor violations
get intimately related to fermion masses and neutrino
oscillations.

5) $\bf{A-Terms\ Induced\ Through\ RG\ Running\ from\ M^{*}\ to\
M_{GUT}}$: Even if $A_o$ is zero at $M^{*}$ (as we assume, for
concreteness, see also \cite{gauginomed} and \cite{faraggiJCP}),
RG running from $M^{*}$ to $M_{GUT}$ in the context of
SO(10)/G(224) would still induce non-zero $A$ parameters at the
GUT scale \cite{BarbieriHallStrumia}. For our case, the $A$ terms
are induced through loop diagrams involving the $h_{33}$,
$g_{23}$, and $a_{23}$ couplings and the SO(10) or G(224)
gauginos. We find that if we take $M_{10_H}\approx\ M_{16_H} \
\approx\ M_{GUT}$, we can write the $A_{LR}$-matrix at the
GUT-scale for the down squark sector in the SUSY basis for the
case of SO(10) as follows:
\begin{eqnarray}
\label{eq:Ad} A^d_{LR} =Z\times\ (X^d_L)^\dagger\left[
\begin{array}{ccc}
0&95\epsilon'+90\eta' &0\\-95\epsilon'+90\eta'
&63\zeta_{22}^d&95\epsilon+90\eta-27\sigma\\0&
-95\epsilon+90\eta-27\sigma&63
\end{array}\right] X^d_R
\end{eqnarray}

\noindent where $Z = \bigl(\frac{1}{16\pi^2}\bigr) h_t g_{10}^2
M_{\lambda} ln\bigl(\frac{M^{*}}{M_{GUT}}\bigr)$, and
$(X^d)_{L,R}$ are given in Eqs. (\ref{eq:xdl}) and (\ref{eq:xdr}).
The $g_{23}$ coupling does not contribute to the up-sector; thus
the $A$-matrix for the up squarks, $A^u_{LR}$, can be obtained
from above by setting $\eta' = 0$ and replacing $90\eta-27\sigma$
by $63\sigma$, $\zeta_{22}^d$ by $\zeta_{22}^u$, and $X^d_{L,R}$
by $X^u_{L,R}$ in $A^d $. Similarly, the lepton $A$-matrix,
$A^l_{LR}$ is obtained by letting $(\epsilon,\
\epsilon')\rightarrow -3(\epsilon,\ \epsilon')$ and replacing
$X^d_{L,R}$ by $X^l_{L,R}$ in $A^d_{LR}$. For the case of G(224),
the matrix $A^d_{LR}$ would be obtained by making the
substitutions: (90, 63, 95) $\rightarrow$ (42, 27, 43) in Eq.
(\ref{eq:Ad}), and likewise in $A^u_{LR}$ and $A^l_{LR}$. It is
sometimes convenient to define the sfermion transition mixing
angles as
\begin{eqnarray}
\label{eq:deltalr} (\delta^{d,l}_{LR})_{ij}\ \equiv\
(A^{d,l}_{LR})_{ij}\ \bigl(\frac{v_d}{m_{sq}^2}\bigr)\ =\
(A^{d,l}_{LR})_{ij}\ \bigl(\frac{v_u}{\tan\beta\
m_{sq}^2}\bigr);\quad\ (\delta^{u}_{LR})_{ij}\ \equiv\
(A^{u}_{LR})_{ij}\ \bigl(\frac{v_u}{m_{sq}^2}\bigr)~.
\end{eqnarray}

Note that these induced $A_{LR}$-terms for all three sectors, like
the squarks $(mass)^2$ elements $\delta_{LL,RR}^{ij}$ given in
Eqs. (\ref{eq:deltaij})-(\ref{eq:deltallrr}), are completely
determined within our model by the fermion mass matrices, for a
given choice of $M_{\lambda}\approx m_{1/2}$ and
$ln(M^{*}/M_{GUT})$. We now utilize these SUSY CP and
flavor-violating elements to predict the results of our model.

Once again, as in the case of $\hat{\delta}^{ij}_{LL,RR}$, these
induced $A$-terms arising purely through GUT-physics, would be
absent or negligibly small in CMSSM. As a result, some of the
interesting predictions of our model as regards $\epsilon'_K$ and
edm's (to be discussed below) and lepton flavor violations (to be
discussed in a forthcoming paper \cite{LFV}) would be absent
altogether in CMSSM.

\section{Compatibility of CP and Flavor Violations
with Fermion Masses and Neutrino Oscillations in SO(10)/G(224):
Our Results}

It has been noted in Sec. 1 that (given about 15\% uncertainty in
the matrix elements) the SM agrees very well with all four entries
of Eq. (\ref{eq:4qtty}), for a single choice of the Wolfenstein
parameters $\bar\rho_W$ and $\bar\eta_W$ (Eq. (\ref{eq:rhoeta})).
The question then arises (as noted in Sec. 1): If a SUSY SO(10) or
G(224) model is constrained by requiring that it should
successfully describe fermion masses and neutrino oscillations (as
in Sec. 2), can it still yield (for some choice of phases in the
parameters $\eta,\ \sigma,\ \epsilon$ etc.) values for
$\bar\rho_W$ and $\bar\eta_W$ more or less in accord with the
SM-based phenomenological values for the same, as listed in Eq.
(\ref{eq:rhoeta})? Anticipating that (for any given choice of the
parameters $\eta,\ \sigma,\ \epsilon$ etc.) the SO(10)/G(224)
model-based values of $\bar\rho_W$ and $\bar\eta_W$ would
generically differ from the SM-based phenomenological values
(given in Eq.(\ref{eq:rhoeta})), we will denote the former by
$\bar\rho'_W$ and $\bar\eta'_W$ and the corresponding
contributions from the SM-interactions (based on $\bar\rho'_W$ and
$\bar\eta'_W$) by $SM'$. The question that faces us then is this:
When the $SM'$ contributions are added to the SUSY contributions
arising from the three sources listed in Sec. 4, can such a
constrained SO(10) or G(224) model account for the observed values
of all the four quantities listed in Eq. (\ref{eq:4qtty}), and in
addition is it consistent with the empirical upper limits on the
edm's of the neutron and the electron?$\footnote{We extend the
same question to include lepton flavor violating processes (such
as $\mu\rightarrow e\ \gamma$ and $\tau\rightarrow \mu\ \gamma$)
in a separate note.}.$

Before presenting our results, we make some preliminary remarks.
First of all one might have thought, given the freedom in the
choice of phases in the parameters of the mass matrices, that it
ought to be possible to get almost any set of values of
($\bar\rho_W$ and $\bar\eta_W$), and in particular those in accord
with the SM values (Eq. (\ref{eq:rhoeta})). It turns out, however,
that in general this is indeed not possible without running into a
conflict with the fermion masses and/or neutrino oscillation
parameters within a SO(10) or G(224)-model\footnote{for a
discussion of difficulties in this regard within a recently
proposed SO(10)model, see e.g. \cite{GohMohapatNg}}. In other
words, any predictive SO(10) or G(224)-model is rather constrained
in this regard.

Second, one might think that even if the SO(10)/G(224)
model-derived entities $\bar\rho'_W$ and $\bar\eta'_W$,
constrained by the pattern of fermion masses and neutrino
oscillations, are found to be very different in signs and/or
magnitudes from the SM values shown in Eq. (\ref{eq:rhoeta}),
perhaps the SUSY contributions added to the $SM'$
contributions(based on $\bar\rho'_W$ and $\bar\eta'_W$) could
possibly account for all four quantities listed in Eq.
(\ref{eq:4qtty}). {\it It seems to us, however, that this is
simply not a viable and natural possibility}, unless one is
willing to invoke MSSM and finely adjust its arbitrary (in general
some 105) parameters, as needed. In the latter case, the good
agreement between experiments and the SM predictions would appear
to be fortuitous (see Sec. 1).

This is why it seems to us that the only viable and natural
solution for any SUSY G(224) or SO(10) model for fermion masses
and neutrino oscillations is that the model, allowing for phases
in the fermion mass matrices, should not only yield the masses and
mixings of all fermions including neutrinos in accord with
observations (as in Sec. 2), but it should yield $\bar\rho'_W$ and
$\bar\eta'_W$ that are close to the SM values shown in Eq.
(\ref{eq:rhoeta}). This, if achievable, would be a major step in
the right direction. One then needs to ask: how does the combined
($SM'$ + SUSY) contributions fare for such a solution as regards
its predictions for the four quantities of Eq. (\ref{eq:4qtty})
and other CP and/or flavor violating processes, for any given
choice of the SUSY parameters ($m_o,\ m_{1/2},\ A_o,\ \tan\beta$
and $sgn(\mu)$)? It should be stressed here that even if the CKM
elements including $\bar\rho'_W$ and $\bar\eta'_W$ should turn out
to be close to the SM values (Eq. (\ref{eq:rhoeta})), {\it the
SUSY contributions can in general still have a marked effect, in
accord with observations, at least on some of the processes where
the SM (or $SM'$)-contributions are naturally suppressed} (as in
the case for $\epsilon_K$, edm's and lepton flavor violating
transitions). Study of these processes, some of which we discuss
below, can help distinguish between the SM versus the SUSY
SO(10)/G(224)-models.

Without further elaboration, we now present our main results. In
this paper we will present only one fit to the parameters which
has the desired properties.$\footnote{ We have verified that there
actually exists a class of fits which nearly serve the same
purpose. Only one of these (Eq. (\ref{eq:fitA})) is exhibited here
for the sake of concreteness.}$
Allowing for phases ($\sim$ 1/10 to 1/2) in the parameters $\eta,\
\sigma,\ \epsilon',\ and\ \zeta_{22}^d$ of the G(224)/SO(10)
framework (see Eq. (\ref{eq:fit})) we find that there do exist
solutions which yield masses and mixings of quarks and leptons, in
accord with observations to within 10\% for most part (see
discussion below), and at the same time yield $\bar\rho'_W$ and
$\bar\eta'_W$ close to the SM values, as given in Eq.
(\ref{eq:rhoeta}). A desired fit to the parameters is given by:
\begin{eqnarray}
\label{eq:fitA}
\begin{array}{l}
\ \ \sigma = 0.109-0.012i, \quad \eta = 0.122-0.0464i, \quad
\epsilon =
-0.103, \quad \eta' = 2.4 \times 10^{-3},\\
\begin{array}{l}
\ \epsilon' = 2.35\times 10^{-4}e^{i(69^{\circ})},\quad
\zeta_{22}^d = 9.8\times 10^{-3}e^{-i(149^{\circ})},\quad({\cal
M}_u^0,\ {\cal M}^0_d)\approx (100,\ 1.1) \mbox{ GeV}.
\end{array}
\end{array}
\end{eqnarray}

For the sake of simplicity and economy, we have set $\zeta_{22}^u
= 0$ in this fit; however, values of $|\zeta_{22}^u |\lsim
10^{-3}$ can lead to similar results. Note that the magnitudes of
the real parts of $\eta,\ \sigma,\ \epsilon,$ and $\epsilon'$ are
nearly the same as those given in the CP-conserving case
\cite{BPW} (see Eq. (\ref{eq:fit})); in particular the relative
signs of these real parts are identical.  The fit shown above
leads to the following values for the fermion masses and mixings,
while preserving the predictions for the neutrino system as in Eq.
(\ref{eq:pred}):
\begin{eqnarray}
\label{eq:predA}
\begin{array}{l}
\ \ \ \ (m_t^{\rm phys},\ m_b(m_b),\ m_{\tau})\approx (174,\
4.97,\ 1.78)\
GeV\\
\begin{array}{l}
\ \ \ (m_c(m_c),\ m_s(1GeV),\ m_{\mu})\approx (1320,\ 101,\ 109)\
MeV\\
\begin{array}{l}
\ \ (m_u^{\circ}(1GeV),\ m_d^{\circ}(1GeV),\ m_e^{\circ})\approx
(10.1,\ 3.7,\ 0.13)\ MeV\\
\begin{array}{l}
\ (V_{us},\ V_{cb},\ |V_{ub}|,\ |V_{td}|)(\le m_Z)\approx
(0.2250,\ 0.0412,\ 0.0037,\ 0.0086)\\
\begin{array}{l}
\bar\rho'_W = 0.150,\quad \bar\eta'_W = 0.374
\end{array}
\end{array}
\end{array}
\end{array}
\end{array}
\end{eqnarray}

In obtaining the fermion masses at the low scales, we have not
directly used ${\cal M}_u^0$ and ${\cal M}_d^0$ of Eq.
(\ref{eq:mat}). Instead, we have used: (a) $m_t(m_t)=167\ GeV$ and
$m_{\tau}(m_{\tau})=1.777\ GeV$ as inputs; (b) the GUT-scale
predictions of our model for the {\it ratios} of masses - such as
$m_b/m_{\tau},\ m_{u,c}/m_t,\ m_{d,s}/m_b,\ m_{\mu}/m_{\tau}$ etc;
(c) renormalization in 2-loop QCD of these ratios in going from
the GUT-scale to an effective SUSY-scale $M_S=500\ GeV$; and (d)
the evolutions in 3-loop QCD and 1-loop QED of individual fermion
masses as the effective momentum runs from $M_S$ to the
appropriate low energy scales \cite{reneq}$\footnote{Defining
$\eta_{a/b}\equiv (m_a/m_b)_{GUT}/(m_a/m_b)_{M_S}$ and
$\eta_f\equiv m_f(M_S)/m_f(\mu_{low})$, we get (for tan$\beta$ = 5
and $\alpha_3(M_Z)$=0.118): $\eta_{b/\tau}=0.6430,\
\eta_{u,c/t}=0.4456,\ \eta_{d,s/b}=0.7660,\
\eta_{e,\mu/\tau}=0.9999,\ \eta_u=0.3954,\ \eta_{d,s}=0.3982,\
\eta_{c}=0.4418,\ \eta_{b}=0.6053,\ \eta_{e,\mu}=0.9894,\
\eta_{\tau}=0.9914,\ \eta_t=0.9427$. The CKM elements at low
scales are given by $V_{\alpha\beta}(\le m_Z)=
V_{\alpha\beta}(GUT)/K_{\alpha\beta}$, where
$K_{\alpha\beta}\approx 0.91$ for $\alpha\beta = ub,\ cb,\ td,\
and\ ts$ and $K_{\alpha\beta}\approx 1$ for the other elements.}$.

The primes on $\bar\rho'_W$ and $\bar\eta'_W$ signify that these
values are obtained from the SO(10)/G(224) model based fermion
mass matrices (as in Eq. (\ref{eq:mat})), {\it in conjunction
with} fermion masses and neutrino oscillations, as opposed to
SM-based phenomenological values (Eq. (\ref{eq:rhoeta})).

Note that, except for the very light fermion masses
$(m_u^{\circ},\ m_d^{\circ},$ and $m_e^{\circ})$ which would need
corrections of order 1 to few MeV \cite{a}, all the other
quark-lepton masses and especially the CKM mixings are in good
accord with observations (see values quoted below Eq.
(\ref{eq:rhoeta}) or Ref. \cite{b}), to within 10\%. (As alluded
to before, we should not of course expect the very light fermion
masses to be described adequately by the gross pattern of the
mass-matrices exhibited in Sec. 2. In particular the ``11''
entries in Eq. (\ref{eq:mat}) (expected to be of order
$10^{-4}-10^{-5}$) arising from higher dimensional operators,
which have been neglected in Sec. 2, can quite plausibly lead to a
needed reduction in $m_u$ by about 6-8 MeV and an increase in
$(m_e$ and $m_d)$ $\footnote{Note that the ``11'' entry for the up
sector can differ from that for the down sector even in sign
because of contribution through the operator ${\bf 16}_1{\bf 16}_1
{\bf 16}_H^d({\bf 16}_H/M'')(S/M)^{n}$ which contributes only to
$m_d\ and\ m_e$ (so that $\delta m_d \ =\ \delta m_e$ at
$M_{GUT}$) but not to $m_u$.}$ by nearly (0.36 and 2-3) MeV
respectively, at the 1 GeV scale, $\it{without}$ altering the CKM
mixings).

The important point is that the SO(10)/G(224)-model presented in
Sec. 3 has turned out to be capable of yielding values for
$\bar\rho'_W$ and $\bar\eta'_W$ that are close to the SM values as
desired, while simultaneously being able to yield fermion masses
of the two heavy families, all the CKM elements and neutrino
oscillations (see Sec. 2), in good accord with observations. This
in itself is non-trivial.

Before presenting the results for CP and flavor violations some
comments are in order as regards the parameters of the model
versus its predictiveness. As expected, introduction of (in
general four) phases in the Dirac mass matrices clearly increase
the number of parameters compared to that for the CP-conserving
case \cite{BPW}. As a result, as long as we confine to the realm
consisting of (a) the fermion masses and mixings, (b) CP and
flavor violations induced $\it{only\ by\ the\ SM\ interactions}$,
and (c) neutrino oscillations, the predictiveness of the model is
reduced considerably (compare with the CP-conserving case of Ref.
\cite{BPW}, see Sec. 2), the number of parameters now being
comparable to the number of observables. Nevertheless, some gross
features of the predictions in fact survive, even in the realm
mentioned above, simply because: (a) the entries in the
mass-matrices, governed by flavor symmetries, are hierarchical
with a pattern as in Eq. (\ref{eq:mat}); (b) the phases are
constrained$\footnote{For instance, consider the familiar relation
$|V_{us}| = |\sqrt{m_d/m_s}-e^{i\phi}\sqrt{m_u/m_c}|$, that holds
for a hierarchical pattern. Given $\sqrt{m_d/m_s}\approx 0.22$ and
$\sqrt{m_u/m_c}\approx 0.07$, we cannot of course predict $V_{us}$
precisely without knowing the phase angle $\phi$. Yet, since
$\phi$ can vary only between 0 to 2$\pi,\ |V_{us}|$ must lie
between 0.15 and 0.29}$ to lie between 0 to 2$\pi$, and, (c) the
system itself is constrained by the group theory of
SO(10)/SU(4)$^c$. One can argue that these features in turn pretty
much ensure the gross nature of the following predictions: (i)
$m_b^{\circ}/m_{\tau}^{\circ}\approx 1$, (ii) $|V_{us}|\sim 0.2$,
(iii) $|V_{ub}|\approx \sqrt{m_u/m_c}\ |V_{cb}|$, (iv)
$|V_{td}|\approx \sqrt{m_d/m_s}\ |V_{cb}|$, (v)
$m_{\nu_2}/m_{\nu_3}\sim 1/10$, (vi) $m_{\nu_3}\sim 1/10$ eV, and,
(vii) $\sin^2 2\theta_{\nu_{\mu} \nu_{\tau}}^{osc}\approx
(0.8-0.99)$, despite large variations in the parameters.

The real virtue of the model (including the phases) emerges, once
one includes SUSY CP and flavor violations. In this case, the
realm of observables and thereby the predictiveness of the model
expands enormously. The set of observables now includes not only
the four entities listed in Eq. (\ref{eq:4qtty})-i.e., $\Delta
m_K,\ \epsilon_K,\ \Delta m_{B_d}$ and $S(B_d\rightarrow J/\psi
K_S)-$ $\it{but\ also}$ a host of others, for which the
predictions of the G(224)/SO(10) model including ($SM'$+ SUSY)
contributions, can a priori differ significantly from those of the
SM contributions. In particular, the set includes observables such
as (v) $\epsilon'_K$ (vi) $\Delta m_{B_s}$, (vii)
$S(B_d\rightarrow \phi K_S)$, (viii) $S(B_d\rightarrow \eta'
K_S)$, (ix) $S(B_s\rightarrow J/\psi \phi)$, (x) $S(B_s\rightarrow
\phi K_S)$, (xi) $B\rightarrow K\pi$, (xii) $B\rightarrow \pi\pi$
(rates and asymmetry parameters), (xiii) $b\rightarrow s\gamma$,
(xiv) electric dipole moments of ($n, e, Hg, d$) and (xv) Lepton
flavor violating processes ($\mu\rightarrow e\gamma,\
\tau\rightarrow\mu\gamma,\ \tau\rightarrow e\gamma$), and more.

Now, the SUSY contributions do of course depend in part on the
flavor preserving SUSY-parameters (i.e. $m_o,\ m_{1/2},\ \mu$, and
$\tan\beta$; we set $A_o$ = 0 at $M^{*}$). But these few
parameters should be regarded as {\it extraneous} to the present
model, and hopefully, they would be determined through the
discovery of SUSY at the LHC. The interesting point is that for a
given choice of these flavor-preserving SUSY parameters
(essentially $m_o\ and\ m_{1/2}$) the SUSY contributions to
$\it{all}$ the CP and/or flavor-violating processes listed above
get completely determined within our model, $\it{in\ magnitude\
as\ well\ as\ in\ phases}$. This is because all the flavor and in
general CP violating sfermion
$(mass)^2$-parameters$\footnote{counting the number of such
$(mass)^2$-parameters}$ (($\delta m^2)^{ij}_{LL,RR,LR})$, arising
through SO(10)/G(224)-based RG running from $M^{*}$ to $M_{GUT}$
are completely fixed in the model in terms of the parameters of
the fermion mass-matrices (see Eqs. (\ref{eq:deltaij}),
(\ref{eq:Ad}) and (\ref{eq:deltalr})). The latter are, however
essentially fixed by fermion masses and mixings, as shown in the
fit given above (Eq. (\ref{eq:fitA})), especially when we demand
that the $\bar\rho'_W$ and $\bar\eta'_W$ be close to the
SM-values. In short, the inclusion of SUSY CP and flavor
violations, treated in conjunction with fermion masses and
neutrino oscillations, encompasses a host of processes without
introducing new parameters and thereby increases the
predictiveness of the model enormously.

In this paper, we will present the results for some of the
processes listed above, in particular those shown in Eq.
(\ref{eq:4qtty}) as well as those for $\epsilon'_K$ and the edm's
of the neutron and the electron. Some of the other processes
including lepton flavor violation will be considered in a separate
paper.

Using Eqs. (\ref{eq:deltaij}) and
(\ref{eq:deltam'l})-(\ref{eq:deltalr}) for the squarks $(mass)^2$
elements ($\delta_{LL,RR,LR}^{ij}$ etc.) as predicted in our
model, the expressions given in Refs.
\cite{Gabbiani,Ciuchini2,Khalil,Buras} for the SUSY contributions,
and the values of $m_s,\ m_c$ and the CKM elements (including
$\bar\rho'_W$ and $\bar\eta'_W$) as obtained in the fit given
above (see Eqs. (\ref{eq:fitA}) and (\ref{eq:predA})), we can now
derive the values of the four entities listed in Eq.
(\ref{eq:4qtty}), treating separately the cases of the SO(10) and
the G(224)-models. For reasons explained below Eq.
(\ref{eq:deltambr}), the SUSY contributions are reduced (in most
cases) by about a factor of 2.5 in the amplitude for the case of
G(224) compared to that of SO(10), being the effective symmetry in
4D. This distinction, as we will see, provides a way to
distinguish between the SO(10) and the G(224)-models
experimentally. The predictions of the model (corresponding to the
fit shown in Eq. (\ref{eq:fitA})) are shown in table 1. We have
included both the $SM'$ and the SUSY contributions in obtaining
the total contributions (denoted by ``Tot''). In quoting the
numbers we have fixed, for concreteness, $M^{*}/M_{GUT}\approx$ 3
and thus $\xi\approx$ 0.4, and have made a plausible choice for
the SUSY spectrum - i.e. $m_{sq}\approx$ (0.8 - 1) TeV with
$x=m_{\tilde g}^2/m_{sq}^2 \approx$ 0.8, although a variation in
these parameters with $m_{sq}$ as low as about 600 GeV or $x = 0.5
- 0.6$ can still lead to the desired results for all four
quantities especially for the case of G(224) (see remarks below).

\vspace*{12pt}

\noindent
\begin{tabular}{|c|c|c|c|c|}
\hline \rule[-3mm]{0mm}{8mm} ($m_o,\ m_{1/2}$)(GeV) &
\multicolumn{2}{c|}{(800, 250)}&\multicolumn{2}{c|} {(600, 300)}\\
\hline
& (a)& (b) & (c) & (d) \\
 & SO(10) & G(224) & SO(10) & G(224)  \\
\hline

$\Delta m_K^{s.d.}$(Tot $\approx\ SM'$)(GeV) & $2.9\times10^{-15}$
& 2.9$\times 10^{-15}$
         & 2.9$\times 10^{-15}$  & 2.9$\times10^{-15}$
                  \\ \hline
$\epsilon_K (SM')$   & 2.83$\times 10^{-3}$  & 2.83$\times
10^{-3}$
         & 2.83$\times 10^{-3}$  & 2.83$\times 10^{-3}$
                 \\ \hline
$\epsilon_K (Tot)$   & 1.30$\times 10^{-3}$  & 2.32$\times
10^{-3}$
         & 2.01$\times 10^{-3}$  & 2.56$\times 10^{-3}$
          \\ \hline
$\Delta m_{B_d}$ (Tot $\approx\ SM'$)(GeV)  & 3.62$\times
10^{-13}$ & 3.56$\times 10^{-13}$
         & 3.58$\times 10^{-13}$  & 3.55$\times 10^{-13}$
          \\ \hline
$S(B_d\rightarrow J/\psi K_S)$ (Tot $\approx\ SM'$)  & 0.740 &
0.728
         & 0.732 & 0.726
          \\ \hline
\end{tabular}

\vskip.10in Table 1. {\small Predictions of the SUSY SO(10) and
G(224) models corresponding to the fit for the fermion
mass-parameters shown in Eq. (\ref{eq:fitA}). Either model with
the fit as in Eq. (\ref{eq:fitA}) leads to the fermion masses and
CKM mixings in good agreement with the data (see Eq.
(\ref{eq:predA})). The total contribution (denoted by ``Tot'')
represents the sum of the $SM'$ and the SUSY contributions. Note
that the SUSY contribution is important only for $\epsilon_K$,
furthermore it is relatively negative (as desired) compared to the
$SM'$ contribution (see discussion in text). The superscript
$s.d.$ on $\Delta m_K$ represents short distance contribution.}
\vspace*{12pt}

In obtaining the entries for the $K$-system we have used central
values of the matrix element $\hat{B}_K$ and the loop functions
$\eta_i$ (see Refs. \cite{Buras, Ciuchinietal} for definitions and
values) characterizing short distance QCD effects  - i.e.
$\hat{B}_K = 0.86\pm 0.13,\ f_K = 159\ MeV,\ \eta_1 = 1.38 \pm
0.20$,$\footnote{We will be guided by the error of $\pm 0.20$ on
$\eta_1$, used in \cite{Buras}, although that quoted in
\cite{Ciuchinietal} is considerably larger ($\pm$ 0.53).}$ $\eta_2
= 0.57 \pm 0.01$ and $\eta_3 = 0.47 \pm 0.04$. For the $B$-system
we use the central values of the unquenched lattice results:
$f_{B_d}\sqrt{\hat{B}_{B_d}} = 215(11)(^{+0}_{-23})(15)\ MeV$ and
$f_{B_s}\sqrt{\hat{B}_{B_s}} = 245(10)(^{+3}_{-2})(^{+7}_{-0})\
MeV$ \cite{SAoki}. Note that the uncertainties in some of these
hadronic parameters are in the range of 15\%; thus the predictions
of our model as well as that of the SM would be uncertain at
present to the same extent.

At this stage the following comments are in order.

(1) In the cases of $\Delta m_K,\ \Delta m_{B_d}\ and\
S(B_d\rightarrow J/\psi K_S)$, the SUSY contributions (with
$m_{sq}\sim$ 0.8-1 TeV and $x \sim$ 0.5-0.8) are found to be
rather small ($\sim$ 0.5\%, 2\%, and 3\% respectively) compared to
the $SM'$ contribution. As a result, for these three entities, the
$SM'$ contribution practically coincides with the total
contribution, which is what is shown in the table. {\it By
contrast, for the same spectrum, the SUSY-contribution to
$\epsilon_K$ is found to be rather sizeable ($\sim$
20-25\%)$\footnote{The fact that the SUSY contribution to
$\epsilon_K$ (in contrast to those for $\Delta m_K,\ \Delta
m_{B_d}$ and $S(B_d\rightarrow J/\psi K_S)$) is relatively large
is simply because the SM contribution to $\epsilon_K$ is strongly
suppressed owing to the smallness of the relevant CKM mixings.}$,
and importantly enough, negative compared to the
$SM'$-contribution}$\footnote{In as much as we require
$\bar\rho'_W$ and $\bar\eta'_W$ to be close to the SM-based
phenomenological values (as in Eq. (\ref{eq:rhoeta})), in accord
with the observed values of the fermion masses, CKM-elements and
neutrino oscillation parameters, we find that the class of fits
satisfying these requirements lead to SUSY-contribution to
$\epsilon_K$ that is relatively negative compared to the
$SM'$-contribution.}$. The fact that it is relatively negative is
an outcome of the model and, as it turns out, is most desirable
(see below).

(2) Comparing the predicted values shown in Table 1 with the
observed ones (see those listed below Eq. (\ref{eq:rhoeta})),
together with $\Delta m_K^{obs} = 3.47\times 10^{-15}\ GeV$, we
see that all four entities including $\epsilon_K$ and the
asymmetry parameter $S(B_d\rightarrow J/\psi K_S)$ agree with the
data quite well, for the cases of SO(10) as well as G(224) shown
in the last two columns (i.e. for $m_{sq}\approx$ 1 TeV, and $x
\approx$ 0.8), and also for the case of G(224) in the second
column ($m_{sq}\approx$ 800 GeV, $x \approx$ 0.8). In making this
comparison we are allowing for plausible (at present theoretically
uncertain but allowed) long distance contribution to $\Delta m_K
(\sim \pm 15\%)$, and uncertainties in $\hat B_K\ or\ \eta_1 \lsim
10\%$ (see entries for $\epsilon_K$ in the last three columns) and
that in $f_{B_d}\sqrt{\hat{B}_{B_d}}$ by about 3\%.

(3) We note that a choice of the SUSY-parameters, e.g. ($m_o,\
m_{1/2}$) = (800, 250) GeV, shown in the table, is in accord with
the WMAP-constraint on CMSSM-spectrum in the event that the
lightest neutralino is the LSP and represents cold dark matter
\cite{JEllis}.

(4) It is crucial  that the SUSY contribution to $\epsilon_K$ (as
mentioned above) is significant and is negative relative to the
$SM'$-contribution. Indeed this is what makes it possible for
$\epsilon_K (Tot)$ to be desirably lower than the $\epsilon_K
(SM')=2.83\times 10^{-3}$ and thereby to agree better in the last
three columns (cases b, c and d) with $\epsilon_K^{obs} =
2.27\times 10^{-3}$. Had the SUSY contribution been positive
relative to the $SM'$ contribution and still as significant as
above, $\epsilon_K (Tot)$ would have been (3.34, 3.53, and
3.10)$\times 10^{-3}$ for the cases (b), (c), and (d)
respectively, in strong disagreement with observation. {\it In
short, the SUSY contribution of the model to $\epsilon_K$ has just
the right sign and nearly the right magnitude} to play the desired
role. This seems to be an intriguing feature of the model.

(5) Since the values of the CKM elements including $\bar\rho'_W$
and $\bar\eta'_W$ obtained within our model (see Eq.
(\ref{eq:predA})) are quite close to the SM based phenomenological
values (see Ref. \cite{Ciuchinietal} and Eq. (\ref{eq:rhoeta})),
we would of course expect that the $SM'$ contributions should
nearly be the same as the SM contributions, for the same choice of
the hadronic parameters ($\hat{B}_K,\ \eta_i,\
f_{B_d}\sqrt{\hat{B}_{B_d}}$ etc.). For instance, using the
central values of the parameters given in the recent update of the
CKM-triangle analysis by M. Bona et al. \cite{ASoni}, that is,
$\lambda= |V_{us}|=0.2265,\ |V_{cb}| = 4.14\times 10^{-2},\
\bar{\rho}_W = 0.172,\ \bar{\eta}_W = 0.348,\ m_c = 1.3\ GeV\ and\
f_K=159\ MeV$, and the hadronic parameters as in our case - that
is, $\hat{B}_K = 0.86,\ \eta_1 = 1.38,\ \eta_2 = 0.57$, and
$\eta_3 = 0.47$ - one obtains $\epsilon_K(SM)=2.72\times 10^{-3}$
which is about 20\% higher than the observed value. Contrast this
with the predictions for $\epsilon_K$(Tot) of the SO(10) or G(224)
models for the cases (b), (c) and (d) in Table 1 where the
discrepancies between the predicted and observed value of
$\epsilon_K$ range from 2 to 12\% with varying signs. At present,
such discrepancies, even as high as 20\% for the SM, can of course
be accommodated by allowing for uncertainties in $\hat{B}_K,\
\eta_1$, and also in $\lambda$.

(6) One main point we wish to stress here, however, is this: At
present, the distinctions between the predictions of the SM (in
particular for $\epsilon_K$) versus those of the SUSY SO(10) or
G(224) models on the one hand, and those between the predictions
of the SUSY SO(10) versus the G(224) models on the other hand
(compare columns (a), (b), (c) and (d) of Table 1) are marred in
part because of uncertainties ($\sim 15\%$) in the hadronic
parameters ($\hat{B}_K,\ \eta_1$ etc.) as well as that ($\sim
2\%$) in $\lambda$, and in part because SUSY is not discovered as
yet, and thus the SUSY spectrum is unknown. But once (hopefully)
SUSY is discovered at the LHC and thereby the SUSY parameters get
fixed, and in addition the uncertainties in the hadronic
parameters are reduced (hopefully) to a few percent level through
improvements in the lattice calculations, we see from the analysis
presented above that we can utilize the combined set of
informations to distinguish experimentally between the SM versus
the SUSY SO(10)-model versus the SUSY G(224)-model. It is
intriguing to see that {\it even low energy experiments involving
CP and flavor violations can help distinguish between the SO(10)
versus the G(224) models}, both of which nearly coincide as
regards their predictions for fermion masses and neutrino
oscillations. In this way they can shed light on physics at the
superheavy scale $M^{*}\gsim M_{GUT}$. The experimental
distinctions will of course be even sharper once we include
predictions for the other processes, some of which are presented
below.

(7) {\bf $B_d\rightarrow \phi K_S,\ \Delta m_{B_s}$ and
$b\rightarrow s\gamma$:} We now consider the CP violating
asymmetry parameter $S(B_d\rightarrow \phi K_S)$. For a
representative choice of ($m_o,\ m_{1/2}$) = (600, 300) GeV, we
get $\delta_{LL}^{23}=(1.40-0.012i)\times 10^{-2},\
\delta_{RR}^{23}=-(5.39+6.27i)\times 10^{-3},\
\delta_{LR}^{23}=-(0.29+3.08i)\times 10^{-4}/\tan\beta$ and
$\delta_{RL}^{23}=-(1.92+2.70i)\times 10^{-4}/\tan\beta$ as
predictions of our model (see Eqs. (\ref{eq:deltaij}) and
(\ref{eq:deltalr})). It is easy to verify that the SUSY-amplitude
for this decay in our model is only of order 1\% (or less)
\cite{phiKs} compared to that in the SM. As a result, adding $SM'$
and SUSY contributions to the decay amplitudes, we obtain:
\begin{eqnarray}
S(B_d\rightarrow\phi K_S)(Tot\approx SM')\approx 0.728
\end{eqnarray}

Allowing for variant fits which also give fermion masses and CKM
mixings in good agreement with observations, we find that
$S(B_d\rightarrow\phi K_S)$ should lie in the range of $\approx$
+0.65 to +0.73. {\it Thus our model predicts that
$S(B_d\rightarrow\phi K_S)$ is close to the SM prediction
($\approx 0.70 \pm 0.10$) and certainly not negative in
sign}.$\footnote{Our prediction in this regard was reported at the
Fujihara seminar \cite{JCPKeK}, held February 23-25, 2004.}$ When
we started writing this paper, BaBar and BELLE data were yielding
widely differing values of $(0.45\pm 0.43\pm 0.07)$ and $(-0.96\pm
0.50^{+0.09}_{-0.07})$ respectively for $S(B_d\rightarrow\phi
K_S)$ \cite{BabarBelleNew}. Most recently, the two groups reported
new values for the asymmetry parameter $S(B_d\rightarrow\phi
K_S)=(+0.50\pm 0.25^{+0.07}_{-0.04})_{BaBar}; (+0.06\pm 0.33\pm
0.09)_{BELLE}$ \cite{BabarBelleNew}, at the Beijing International
Conference on High Energy Physics . Meanwhile there have been many
theoretical and phenomenological attempts \cite{phiKs, Y} to
obtain possible large deviations in $S(B_d\rightarrow\phi K_S)$
from the SM-value, including, in some cases, negative values for
the same (as suggested by the earlier BELLE data). It will thus be
extremely interesting from the viewpoint of the
G(224)/SO(10)-framework presented here to see whether the true
value of $S(B_d\rightarrow\phi K_S)$ will turn out to be close to
the SM-prediction or not.

Including contributions from $\delta_{LL,RR}^{23}$ and
$\delta_{LR,RL}^{23}$ (as predicted in our model), we get:
\begin{eqnarray}
\Delta m_{B_s}(Tot\approx SM')\approx 17.3\ ps^{-1}
\bigl(\frac{f_{B_s}\sqrt{\hat{B}_{B_s}}}{245 MeV}\bigr)^2~.
\end{eqnarray}

\noindent This is of course compatible with the present lower
limit on $\Delta m_{B_s}\gsim 14.4 ps^{-1}$ \cite{ASoni}.

Using $\delta_{RL}^{23}$ given above, we obtain $A(b_L\rightarrow
s_R\gamma)_{\tilde{g}}\approx(1-1.5)\times 10^{-10}\
GeV^{-1}/\tan\beta$. Even allowing for variant fits, the
SUSY-amplitude, in this case, is found to be only about (1.5-5)\%
of the SM amplitude. The same conclusion holds also for
$A(b_R\rightarrow s_L\gamma)_{\tilde{g}}$. In short, our results
for $(B_d\rightarrow \phi K_S),\ \Delta m_{B_s}$ and $b\rightarrow
s\gamma$ nearly coincide with those of the SM.

(8) {\bf Contribution of the A term to $\epsilon'_K$}: Direct CP
violation in $K_L\rightarrow\pi\pi$ receives a new contribution
from the chromomagnetic operator $Q^-_g = (g/16\pi^2) (\bar s_L
\sigma^{\mu\nu} t^a d_R - \bar s_R \sigma^{\mu\nu} t^a
d_L)G^a_{\mu\nu}$, which is induced by the gluino penguin diagram.
This contribution is proportional to $X_{21}\equiv
Im[(\delta^d_{LR})_{21}-(\delta^d_{LR})_{12}^{*}]$, which is
predictable in our model (see Eqs. (\ref{eq:Ad}) and
(\ref{eq:deltalr})). Following Refs. \cite{Buras2} and \cite{Nir},
one obtains:
\begin{eqnarray}
\label{eq:epsilon'} Re(\epsilon'/\epsilon)_{\tilde{g}}\approx 91\
B_G \bigl(\frac{110~ MeV}{m_s + m_d}\bigr) \bigl(\frac{500~
GeV}{m_{\tilde{g}}}\bigr) X_{21}
\end{eqnarray}
where $B_G$ is the relevant hadronic matrix element.
Model-dependent considerations (allowing for $m_K^2/m_{\pi}^2$
corrections) indicate that $B_G \approx $ 1-4, and that it is
positive \cite{Buras2}. Using the prediction of our model (via
Eqs. (\ref{eq:Ad}) and (\ref{eq:deltalr})), for a typical SUSY-
spectrum used in previous considerations (e.g. $(m_o,\ m_{1/2}) =
(600,\ 300)\ GeV$), we obtain: $X_{21}\approx 2.1\times
10^{-5}/\tan\beta$. Note that the sign of $X_{21}$, as derived in
the model, is positive. Inserting this in Eq. (\ref{eq:epsilon'}),
and putting $(m_s+m_d)\approx$ 110 MeV, we get:
\begin{eqnarray}
Re(\epsilon'/\epsilon)_{\tilde{g}}\approx +(8.8\times
10^{-4})(B_G/4)(5/\tan\beta)~.
\end{eqnarray}
We see that if the positive sign of $B_G$ is confirmed by reliable
lattice calculations, the gluino penguin contribution in our model
can quite plausibly give a significant positive contribution to
$Re(\epsilon'/\epsilon)_{\tilde{g}}\approx (4-14)\times 10^{-4}$,
depending upon $B_G\approx 2-4$ and tan$\beta\approx (3-10)$. At
present the status of SM contribution to $Re(\epsilon'/\epsilon)$
is rather uncertain. For instance, the results of Ref.
\cite{TBlum} and \cite{Noaki} based on quenched lattice
calculations in the lowest order chiral perturbation theory
suggest negative central values for $Re(\epsilon'/\epsilon)$. (To
be specific Ref. \cite{TBlum} yields
$Re(\epsilon'/\epsilon)_{SM}=(-4.0\pm 2.3)\times 10^{-4}$, the
errors being statistical only.) On the other hand, using methods
of partial quenching \cite{Golterman} and staggered fermions,
positive values of $Re(\epsilon'/\epsilon)$ in the range of about
$(3-13)\times 10^{-4}$ are obtained in \cite{Bhattacharya}. In
addition, a recent non-lattice calculation based on
next-to-leading order chiral perturbation theory yields
$Re(\epsilon'/\epsilon)_{SM}=(19\pm 2^{+9}_{-6}\pm 6)\times
10^{-4}$ \cite{Pich}. The systematic errors in these calculations
are at present hard to estimate. The point we wish to note here is
that the SUSY-contribution to $Re(\epsilon'/\epsilon)$, in our
model, is significant, and when the dust settles, following a
reliable calculation of $Re(\epsilon'/\epsilon)$ in the SM, it
would be extremely interesting to check whether the
SUSY-contribution obtained here is playing an important role in
accounting for the observed value given by
$Re(\epsilon'/\epsilon)_{obs} = (17\pm 2)\times 10^{-4}$
\cite{AAlavi} or not.

(9) {\bf EDM of the neutron and the electron}: RG-induced
$A$-terms of the model generate chirality-flipping sfermion mixing
terms $(\delta_{LR}^{d,u,l})_{ij}$, whose magnitudes {\it and}
phases are predictable in the model (see Eqs. (\ref{eq:Ad}) and
(\ref{eq:deltalr})), for a given choice of the universal
SUSY-parameters ($m_o,\ m_{1/2},\ and\ \tan\beta$). These
contribute to the EDM's of the quarks and the electron by
utilizing dominantly the gluino and the neutralino loops
respectively. We will approximate the latter by using the
bino-loop. These contributions are given by (see e.g.
\cite{Nath}):
\begin{eqnarray}
\begin{array}{c}
(d_d, d_u)_{A_{ind}} = (-\frac{2}{9},
\frac{4}{9})\frac{\alpha_s}{\pi}\ e\
\frac{m_{\tilde{g}}}{m_{sq}^2}
f\bigl(\frac{m^2_{\tilde{g}}}{m_{sq}^2}\bigr)
Im(\delta^{d,u}_{LR})_{11}\\
\begin{array}{cc}
(d_e)_{A_{ind}} = -\frac{1}{4\pi}\frac{\alpha_{em }}{cos^2
\theta_W}\ e\ \frac{m_{\tilde{B}}}{m_{\tilde{l}}^2}
f\bigl(\frac{m^2_{\tilde{B}}}{m_{\tilde{l}}^2}\bigr)
Im(\delta^l_{LR})_{11} ~.\\
\end{array}
\end{array}
\end{eqnarray}
For a representative choice  $(m_o,m_{1/2})=(600,\ 300) ~GeV$
(i.e. $m_{sq}$ = 1 TeV, $m_{\tilde{g}}$ = 900 GeV, $m_{\tilde{l}}$
= 636 GeV and $m_{\tilde{B}}$ = 120 GeV ), using Eqs.
(\ref{eq:Ad}) and (\ref{eq:deltalr}), we get:
\begin{eqnarray}
\label{eq:edm} (d_d)_{A_{ind}} = \frac{4.15\times 10^{-26}
}{\tan\beta}\ e cm;\quad (d_u)_{A_{ind}} = (-1.6\times 10^{-26}) \
e cm; \quad (d_e)_{A_{ind}} =\frac{1.1\times 10^{-28}}{\tan\beta}\
e cm~.
\end{eqnarray}
The EDM of the neutron is given by $d_n = \frac{1}{3}(4d_d-d_u)$.
Thus for SO(10), with the choice of $(m_o,\ m_{1/2})$ as above,
we get
\begin{eqnarray}
\label{eq:edmN} (d_n)_{A_{ind}}= (1.6, 1.08)\times 10^{-26} e cm\
{\rm for}\  \tan\beta = (5, 10)~.
\end{eqnarray}

Note that these induced $A$-term contributions are larger for
smaller tan$\beta$. For an alternative choice $(m_o,\ m_{1/2}) =
(800,\ 250)\ GeV$, which as mentioned before is compatible with
the WMAP/CDM-constraint \cite{JEllis}, the predicted EDM of the
neutron and the electron are reduced respectively by about 36\%
and a factor of 3.6. The predictions for the G(224)-model are
smaller than those for the SO(10)-model by nearly a factor of two
in all cases.

We should also note that intrinsic SUSY phases denoted by $\phi_A\
=\ arg(A_o^{*} m_{1/2})$ and $\phi_B\ =\ arg(m_{1/2}\mu B^{*})$,
if present, would make additional contributions to EDM's through
gluino and/or chargino/neutralino loops, which should be added to
the contributions shown above. These contributions have been
widely discussed in the literature (see e.g. \cite{Nath}). As is
well known, with $A_o$ = 0 or small ($\lsim$ 1 GeV) at the scale
$M^{*}$ (as we have chosen, following the examples of Refs.
\cite{gauginomed} and \cite{faraggiJCP}), these contributions are
proportional to $(m_{d,e})\mu\ \tan\beta (\sin\phi_B)$. They would
be typically about 50-300 times larger than the values shown above
(Eqs. (\ref{eq:edm}) and (\ref{eq:edmN})), {\it if} the relevant
intrinsic SUSY phases are nearly unity. This is the familiar SUSY
CP problem. The point of the present paper is that even if the
intrinsic SUSY phases are naturally zero or insignificantly small,
as would be true in a theory where the SUSY CP problem is
naturally solved$\footnote{ A possible solution to the SUSY CP
problem could arise as follows. Assume that {\it CP violation
arises spontaneously, only in the visible sector}, through the
VEV's of fields at the GUT-scale, like those of ${\bf 16}_H,\
\overline{{\bf 16}}_H,\ {\bf 45}_H$ and the singlet $S$. One can
argue that the VEV's of at least some of these can be naturally
complex or purely imaginary, consistent with the minimization of
the potential, even if all parameters in the potential are real.
In this case, intrinsic SUSY phases like those in $m_{1/2}$, $A$
and $B$ are, of course, zero. Now if the $\mu$-term can be
derived, through a satisfactory resolution of the $\mu$-problem,
for example, either by the Giudice-Masiero mechanism \cite{AA}, or
by the ideas suggested in \cite{BB}, or by involving a coupling in
the superpotential of the form \cite{C}: $\kappa {\bf 10}_H {\bf
10}_H N$ + $\lambda N^3$ + ..., where the singlet $N$ is not
allowed to couple to the other fields mentioned above, and
acquires a real VEV of order 1 TeV (as needed), with $\kappa$ and
$\lambda$ being real, then the $\mu$-term would also be real. In
this case, all intrinsic SUSY phases would disappear. We plan to
explore this possibility in a future work.}$$^,$ $\footnote{ An
alternative resolution of the SUSY CP problem arises in a class of
gaugino mediated SUSY-breaking (with the $\mu$-problem being
resolved for example as in \cite{AA}) in which all relevant SUSY
parameters become proportional to $m_{1/2}$ \cite{gauginomed}. A
third resolution of the SUSY CP problem would arise in a model
where both $A$ and $B$ terms are naturally zero or sufficiently
small at the scale $M^{*}$. This is precisely what happens in, for
example, the anomalous $U(1)_A$ model of SUSY-breaking that arises
in the context of a three-family string-solution
\cite{faraggiJCP}. In this case, extra gauge symmetries of the
model suppress both $A$ and $B$ terms at $M^{*}$. }$, {\it the
induced A-term contributions to EDM's (arising from GUT-scale
physics) as presented above would still be present.} The
interesting point is that these contributions are {\it completely
determined} in magnitude and phase within our model (for a given
choice of the SUSY universal parameters $(m_o,\ m_{1/2},\
\tan\beta)$).

Given the experimental limits $d_n<6.3\times 10^{-26}$ e cm
\cite{edmneutron} and $d_e<4.3\times 10^{-27}$ e cm
\cite{edmelectron}, we see that the predictions of the model
(arising only from the induced $A$-term contributions)  especially
for the EDM of the neutron is in an extremely interesting range
suggesting that it should be discovered with an improvement of the
current limit by a factor of about 10.

\section{Summary and Concluding Remarks}

In this paper we have explored the possibility that (a) fermion
masses, (b) neutrino oscillations, (c) CP-non conservation and (d)
flavor violations get intimately linked to each other within
supersymmetric grand unification, especially that based on the
symmetry SO(10) or an effective symmetry G(224)= $SU(2)_L\times
SU(2)_R\times SU(4)^c$. In this context, we extend the framework
proposed previously in \cite{BPW}, which successfully described
fermion masses and neutrino oscillations (see Sec. 2), to include
CP violation. We assume, in the interest of predictiveness, {\it
that CP-violation, arising through the SM as well as SUSY
interactions, has its origin entirely (or primarily) through
phases in the fermion mass matrices.} We also assume that
flavor-blind universal SUSY parameters ($m_o,\ m_{1/2}$ and
tan$\beta$ with $A_o$ being small or real) characterize
SUSY-breaking effects at a high scale $M^{*}\gsim M_{GUT}$. In
this case, {\it all the weak scale CP and/or flavor-violating as
well as flavor-preserving sfermion transition-elements
$\delta_{LL,RR,LR}^{ij}$, and the induced A-parameters, get fully
determined within the model}, in their magnitudes as well as in
phases, simply by the entries in the SO(10)-based fermion
mass-matrices, once the few soft parameters ($m_o,\ m_{1/2}$ and
tan$\beta$) are specified. This is how CP and flavor violations
arising jointly from the SM and SUSY interactions, get intimately
tied to fermion masses and neutrino oscillations, within a
predictive SO(10)/G(224)-framework outlined above. The presence of
GUT-scale physics induces enhanced flavor violation with and
without CP violation, which provides a {\it distinguishing
feature} of the model$\footnote{Even in the case of CMSSM, all the
parameters of MSSM at the electroweak scale (some 105 of them) are
of course also all fully determined in terms of the
SUSY-parameters ($m_o,\ m_{1/2}$ and tan$\beta$) and those
involving the fermion masses and mixings. However, in this case,
as mentioned in Sec. 4, owing to the absence of GUT-scale physics
in the interval $M^{*}\rightarrow M_{GUT}$, the most interesting
effects on the entities considered here (e.g. those on
$\epsilon_K,\ \epsilon'_K$ and the EDM's) would be absent or
negligibly small.}$, relative to CMSSM or MSSM.\footnote{While we
have focussed in this paper on the SO(10)/G(224)-model of Ref.
\cite{BPW}, we note that generically such enhanced flavor and/or
CP violations arising from GUT-scale physics would of course be
present in alternative models of SUSY grand unification \cite{new}
as well, as long as the messenger scale $M^{*}$ lies above
$M_{GUT}$. The detailed predictions and {\it consistency} of any
such model as regards flavor and/or CP violations can however
depend (even sensitively) upon the model, and this is worth
examining.}

As mentioned in Secs. 1 and 5, the framework presented above
faces, however, a prima-facie challenge. Including SM {\it and}
SUSY contributions, the question arises, can the framework
successfully describe the observed features of CP and
flavor-violations including those listed in Eq. (\ref{eq:4qtty}),
while retaining the successes of the CP-preserving framework
\cite{BPW} as regards fermion masses and neutrino oscillations?
Our work here shows that the SUSY SO(10)/G(224)-framework proposed
here, which extends the framework of Ref. \cite{BPW}, indeed meets
this challenge squarely. In the process, it makes several
predictions, only some of which are considered here; these can
eventually help distinguish the framework from other alternatives.

Our results can be summarized as follows:

(1) It is found that, with allowance for phases, there exists a
fit to the parameters of the fermion mass-matrices (Eq.
(\ref{eq:mat})) which successfully describes fermion masses, all
the CKM elements and neutrino oscillations as in Ref. \cite{BPW}
(see Eq. (\ref{eq:predA})), and simultaneously yields the
Wolfenstein parameters ($\bar\rho'_W,\ \bar\eta'_W$) that are
close to the phenomenological SM values (Eq. (\ref{eq:rhoeta})).
The merit of obtaining such values for ($\bar\rho'_W,\
\bar\eta'_W$) in accounting for the data on CP and flavor
violations in quantities such as those listed in Eq.
(\ref{eq:4qtty}) has been stressed in Sec. 5.

(2) With these values of ($\bar\rho'_W,\ \bar\eta'_W$), and a
plausible choice of the SUSY-spectrum$\footnote{Lighter masses for
the SUSY particles like $m_{sq}\approx 600~ GeV$ and
$m_{\tilde{g}}\approx 500 ~GeV$ (say) are allowed for the case of
G(224), though not for SO(10) (see discussion following Table
1).}$ (i.e. $m_{sq}\approx$(600-1000) GeV and
$m_{\tilde{g}}\approx$(500-900) GeV, say, it is found that the
derived values of all four quantities (i.e. $\Delta m_K,\
\epsilon_K,\ \Delta m_{B_d}$ and $S(B_d\rightarrow J/\psi K_S)$
agree with the data quite well (allowing for up to 15\%
uncertainty in hadronic matrix elements), see Table 1.

(3) Although the $SM'$ contributions (for the fit shown in Eq.
(\ref{eq:fitA})) nearly coincide with the SM contributions to all
entities, and SUSY-contributions to entities such as $\Delta m_K,\
\Delta m_{B_d}$ and $S(B_d\rightarrow J/\psi K_S)$ are rather
small ($\lsim$ a few\%), the contributions from SUSY, as a rule,
are nevertheless prominent (of order 20-25\%) especially when the
SM (or $SM'$) contributions are suppressed (for example due to
smallness of the mixing angles). Such is precisely the case for
$\epsilon_K,\ \epsilon'_K$ and the edm's of the neutron and the
electron, (as well as for lepton flavor violating processes to be
discussed in a forthcoming paper \cite{LFV}). It is found that the
SUSY-contribution to $\epsilon_K$ is sizable (of order 20-25\%)
and {\it negative} relative to the $SM'$ contribution, just as
desired, to yield better agreement between the predicted and the
observed value (see Table 1).

(4) The sizable negative contribution of SUSY to $\epsilon_K$ in
our model provides an important tool to help distinguish not only
between the SM versus the SUSY SO(10)/G(224)-models, but also
between the SO(10) and the G(224)-models themselves (see Table 1).
Such distinctions would be possible once (hopefully) SUSY is
discovered at the LHC and thereby the SUSY parameters get fixed,
and in addition the uncertainties in the hadronic parameters
($\hat{B}_K$ and $\eta_1$) are reduced to (say) a 5\% level or
better, through improved lattice calculations.

(5) The model predicts that S($B_d\rightarrow\phi K_S$) should lie
in the range of +(0.65-0.73), i.e. close to the SM-prediction.
Given the present still significant disparities between the BaBar
and BELLE results versus the SM-predictions, it would be
interesting to see where the true value of S($B_d\rightarrow\phi
K_S$) would turn out to lie.

(6) It is interesting that the quantity
$X_{21}=Im[(\delta^d_{LR})_{21} - (\delta^d_{LR})^{*}_{12}]$,
relevant for $\epsilon'_K$, is found to be positive in the model.
If the presently indicated sign of the relevant hadronic matrix
element $B_G$ being positive is confirmed, our model would give a
positive contribution to Re($\epsilon'/\epsilon$) which quite
plausibly can lie in the range of $+(4-14) \times 10^{-4}$. While
this is in the interesting range, its relevance can be assessed
only after the associated matrix elements are determined reliably.

(7) The model predicts that the EDM's of the neutron and the
electron should be discovered with improvements in the current
limits by factors of 10 and 100 respectively. (Intrinsic
SUSY-phases, even if present, would not alter this conclusion as
long as there is no large cancellation between different
contributions.)

(8) It would be most interesting to explore the consequences of
the model, involving SUSY contributions, to other processes such
as $B_s\rightarrow J/\psi \phi$, $B_s\rightarrow\phi K_S$,
$B\rightarrow K\pi$, $B\rightarrow\pi\pi$, $B\rightarrow D K$,
$K_L\rightarrow\pi\nu\bar{\nu}$,
$K^{+}\rightarrow\pi\nu\bar{\nu}$, and especially lepton violating
processes (such as $\mu\rightarrow e\gamma,\
\tau\rightarrow\mu\gamma,\ \tau\rightarrow e\gamma$ etc.). We
stress that the net ($SM'$ + SUSY)-contributions to all these
processes involving CP and/or flavor violations are completely
determined within our model. They do not involve any new
parameters. For this reason, the model turns out to be highly
predictive and thoroughly testable. These processes are under
study; a paper on lepton flavor violation is in preparation.

To conclude, the SUSY SO(10)/G(224) framework, as proposed in Ref.
\cite{BPW} and extended here, subject to the assumption of
universality of SUSY parameters, drastically reduces the
parameters for SUSY-contributions to CP and flavor-violations. In
effect, the extension proposed here ties in fermion masses,
neutrino oscillations, CP and flavor violations within a
predictive and testable framework.

\section{Acknowledgements}

We would like to thank Antonio Masiero, Rabindra Mohapatra, Qaisar
Shafi and Amarjit Soni for helpful discussions and Abolhassan
Jawahery for a helpful correspondence. The work of KSB is
supported in part by the Department of Energy Grant Nos.
DE-FG02-04ER46140, DE-FG02-04ER41306 and by an award from Research
Corporation. That of JCP is supported in part by Department of
Energy Grant No. DE-FG-02-96ER-41015.


\begin{thebibliography}{29}

\bibitem{Ciuchinietal} See e.g. M. Ciuchini, E. Franco, F. Parodi, V. Lubicz, L. Silvestrini, and A.
Stocchi, Talk at ``Workshop on the CKM Unitarity Triangle",
Durham, April 2003, hep-ph/0307195.

\bibitem{ASoni} An extensive analysis appears in the Proceedings of ``The CKM Matrix and the Unitarity Triangle'',
ed. by M. Battaglia, A. J. Buras, P. Gambino and A. Strochhi,
hep-ph/0304132. For a very recent update, see M. Bona et al.,
hep-ph/0408079.

\bibitem{BabarBelle} B. Aubert et al. (BaBar Collaboration), Published in ICHEP 2002, Amsterdam 2002,
481-484, hep-ex/0207042; K. Abe et al. (BELLE Collaboration),
Phys. Rev. {\bf D66} 071102 (2002), hep-ex/0208025.

\bibitem{Yanagida} M. Fukugita and T. Yanagida Phys. Lett.
{\bf B174}, 45 (1986); G. Lazarides and Q. Shafi, Phys. Lett, {\bf
B258}, 305 (1991); M.A. Luty, Phys. Rev. {\bf D45}, 455 (1992). In
the context of the model to be presented here, see J.C. Pati,
Phys. Rev. {\bf D68}, 072002 (2003), hep-ph/0209160.

\bibitem{PS} J.C. Pati and A. Salam, Phys. Rev. Lett.
{\bf 31}, 661 (1973); Phys. Rev. {\bf D10}, 275 (1974).

\bibitem{SO(10)} H. Georgi, in Particles and Fields, Ed. by C.
Carlson (AIP, NY, 1975), p.575; H. Fritzsch and P. Minkowski,
Ann. Phys. {\bf 93}, 193 (1975).

\bibitem{JCPKeK} J.C. Pati, {\it ``Neutrino Masses: Shedding light on Unification and Our Origin"},
Talk given at the Fujihara Seminar, KEK Laboratory, Tsukuba,
Japan, February 23-25, 2004, hep-ph/0407220, to appear in the
proceedings.

\bibitem{seesaw} P. Minkowski, Phys. Lett. {\bf B67}, 421 (1977); M. Gell-Mann, P. Ramond and R. Slansky, in:  {\it
Supergravity}, eds. F. van Nieuwenhuizen and D. Freedman
(Amsterdam, North Holland, 1979) p. 315; T. Yanagida, in:  {\it
Workshop on the Unified Theory and Baryon Number in the
Universe}, eds. O. Sawada and A. Sugamoto (KEK, Tsukuba) 95
(1979); S.L. Glashow, in {\it Quarks and Leptons}, Cargese 1979,
eds. M. Levy et al. (Plenum 1980) p. 707; R. N. Mohapatra and G.
Senjanovic, Phys. Rev. Lett. {\bf 44}, 912 (1980).

\bibitem{BPW} K. S. Babu, J. C. Pati and F. Wilczek, {\it ``Fermion masses, neutrino oscillations, and proton decay in the light of SuperKamiokande"} hep-ph/9812538, Nucl.
Phys. {\bf B566}, 33 (2000).

\bibitem{Hall}   These have been introduced in various forms in the
literature. For a sample, see e.g., C. D. Frogatt and H. B. Nielsen, Nucl.
Phys. {\bf B147}, 277 (1979); L. Hall and H. Murayama, Phys. Rev. Lett.
{\bf 75}, 3985 (1995); P. Binetruy, S. Lavignac and P. Ramond, Nucl. Phys.
{\bf B477}, 353 (1996).
In the string theory context, see e.g., A. Faraggi, Phys. Lett.
{\bf B278}, 131 (1992).

\bibitem{FN4} The zeros in ``11", ``13" and ``31" elements signify that they
are relatively small quantities (specified below). While the ``22"
elements were set to zero in Ref. \cite{BPW}, because they are
meant to be $<$``23" ``32"/``33"$\sim 10^{-2}$ (see below), and
thus unimportant for purposes of Ref. \cite{BPW}, they are
retained here, because such small $\zeta_{22}^u$ and
$\zeta_{22}^d$ [$\sim (1/3)\times 10^{-2}$ (say)] can still be
important for CP violation and leptogenesis.

\bibitem{FN26} For G(224), one can choose the corresponding sub-multiplets --
that is, (1, 1, 15)$_H$, (1, 2, $\bar{4}$)$_H$, (1, 2, 4)$_H$, (2,
2, 1)$_H$ -- together with a singlet $S$, and write a
superpotential analogous to Eq. \eqref{eq:Yuk}.

\bibitem{FN6} If the effective non-renormalizable operator like
${\bf 16}_2{\bf 16}_3{\bf 10}_H{\bf 45}_H/M'$ is induced through exchange
of states with GUT-scale masses involving renormalizable couplings, rather
than through quantum gravity, $M'$ would, however, be of order GUT-scale.
In this case $\langle {\bf 45}_H\rangle/M'\sim 1$, rather than 1/10.

\bibitem{FN7} While ${\bf 16}_H$ has a GUT-scale VEV along the SM singlet, it
turns out that it can also have a VEV of EW scale along the
``$\tilde\nu_L$'' direction due to its mixing  with ${\bf
10}_H^d$, so that the $H_d$ of MSSM is a mixture of  ${\bf
10}_H^d$ and  ${\bf 16}_H^d$. This turns out to be the origin of
non-trivial CKM mixings (see Ref. \cite{BPW}).

\bibitem{FN8}  The flavor charge(s) of ${\bf 45}_H$(${\bf 16}_H$) would
get determined depending upon whether $p$($q$) is one or zero (see below).

\bibitem{JCPErice} J. C. Pati, ``{\it Probing Grand Unification Through Neutrino Oscillations, Leptogenesis and Proton Decay}", hep-ph/0305221,
Proceedings of Erice Summer School, 2002, p.194-236, Ed. by A.
Zichichi, Publ. World Scientific.

\bibitem{Wolfenstein} L. Wolfenstein, Phys. Rev. Lett. {\bf 51}, 1945 (1983).

\bibitem{msugra} A. H. Chamseddine, R. Arnowitt and P. Nath, Phys.
Rev. Lett. {\bf 49}, 970 (1982); R. Barbieri, S. Ferrara and C. A.
Savoy, Phys. Lett. {\bf B119}, 343 (1982); L. J. Hall, J. Lykken
and S. Weinberg, Phys. Rev. {\bf D27}, 2359 (1983); L.
Alvarez-Gaume, J. Polchinski and M. B. Wise, Nucl. Phys. {\bf
B221}, 495 (1983).

\bibitem{gauginomed} Z. Chacko, M. A. Luty, A. E. Nelson and E.
Ponton, JHEP, {\bf 0001}, 003 (2000); D. E. Kaplan, G. D. Kribs
amd M. Schmaltz, Phys. Rev. {\bf D62}, 035010 (2000).

\bibitem{anomU(1)} G. Dvali and A. Pomarol, Phys. Rev. Lett. {\bf
77}, 3728 (1996); P. Binetruy and E. Dudas, Phys. Lett. {\bf
B389}, 503 (1996).

\bibitem{faraggiJCP} A. Faraggi and J.C. Pati, Nucl. Phys. $\bf{B256}$, 526 (1998); hep-ph/9712516v3.

\bibitem{dilaton} N. Arkani-Hamed, M. Dine and S. P. Martin,
hep-ph/9803432, Phys. Lett. {\bf B431}, 329 (1998).

\bibitem{BarbieriHallStrumia} R. Barbieri, L. J. Hall and A.
Strumia, Nucl. Phys. $\bf{B 445}$, 219 (1995); hep-ph/9501334. For
analogous considerations see also L. J. Hall, V. A. Kostelecky and
S. Raby, Nucl. Phys {\bf B267}, 415 (1986); F. Borzumati and A.
Masiero, Phys. Rev. Lett. {\bf 57}, 961 (1986).

\bibitem{carena} M. Carena, M. Olechowski, S. Pokorski, and C. E.
M. Wagner, Nucl. Phys. {\bf B419}, 213 (1994); hep-ph/9311222.


\bibitem{LFV} K. S. Babu, J. C. Pati and P. Rastogi, {\it ``Linking Lepton Flavor Violation with Fermion masses, Neutrino Oscillations and CP Violation in SO(10)/G(224)" }, to appear.

\bibitem{GohMohapatNg} H. S. Goh, R. N. Mohapatra and Siew-Phang
Ng, Phys. Rev. {\bf D68} 115008 (2003), hep-ph/0308197.

\bibitem{reneq} For a partial list of refereces for works along
these lines see e.g. H. Arason et al. Phys. Rev. D $\bf{46}$, 3945
(1992); D. J. Castano, E. J. Pirad, P. Ramond, Phys. Rev.
$\bf{D49}$, 4882 (1994); K. S. Babu, C. N. Leung and J.
Pantaleone, Phys. Lett. $\bf{B 319}$, 191 (1993); H. Fusaoka, Y.
Koide, Phys. Rev. $\bf{D57}$, 3986 (1998); C. R. Das, M. K.
Parida, Eur. Phys. J. $\bf{C20}$, 121 (2001).



\bibitem{a} For very recent improved lattice results on light
quark masses, see C. Aubin et al. (HPQCD, MILC and UKQCD
collaboration), Phys. Rev. {\bf D70} 031504 (2004),
hep-lat/0405022, and C. Aubin et al. (MILC collaboration),
hep-lat/0407028. These papers give: $m_s^{\overline{MS}}$(2 GeV) =
76(0)(3)(7)(0)MeV, 1/2 $(m_u + m_d)^{\overline{MS}}$(2 GeV) =
2.8(0)(1)(3)(0) MeV and $m_u/m_d$ = 0.43(0)(1)(8). Using these
results and extrapolating to 1 GeV, one gets: $m_u (1 ~{\rm
GeV})\approx 2.1\pm 0.5\ MeV,\  m_d (1~ {\rm GeV})\approx 4.9\pm
0.6\ {\rm MeV}$ and $m_s (1~ {\rm GeV})\approx 95\pm 10$ MeV. Note
that these values are considerably smaller than those commonly
used in literature. Somewhat larger values of the masses are
quoted in a paper, also based on improved lattice-calculation, by
M. Goeckeler et al. (QCDSF-UKQCD Collaboration) hep-ph/0409312,
which appeared on the web just before the completion of this
paper.

\bibitem{b} For CKM elements, see e.g. Ref. \cite{Ciuchini et al} or most recent update by M. Bona et al. hep-ph/0408079 in Ref. \cite{ASoni}. The bottom quark mass quoted in C. W. Bauer
et al. (Phys. Rev. {\bf D67}, 054012 (2003)) is: $m_b(m_b) =
4.22\pm 0.09\ GeV$, which is lower than our value given in
Eq.(\ref{eq:predA}) by about 10-12\%. Gluino-loop corrections to
$m_b$ for $\mu<0$ and $\tan\beta = 10$ (say) can, however, reduce
$m_b$ by about 6-7\%.

\bibitem{Gabbiani} F. Gabbiani, E. Gabrielli, A. Masiero, L.
Silvestrini, Nucl. Phys. $\bf{B 477}$, 321 (1996); hep-ph/9604387.

\bibitem{Ciuchini2} M. Ciuchini et al., JHEP 9810:008 (1998),
hep-ph/9808328.

\bibitem{Khalil} S. Khalil and E. Kou, Phys. Rev. {\bf D67}, 055009 (2003);
hep-ph/0212023.

\bibitem{Buras}A. J. Buras, Proceedings of the International School of Subnuclear Physics, Erice, Italy 2000, p 200-337, edited by A. Zichichi, Publ. by World Scientific; hep-ph/0101336.

\bibitem{SAoki} S. Aoki et al., JLQCD Collaboration, Phys. Rev. Lett. {\bf 91}, 212001
(2003), hep-ph/0307039.

\bibitem{JEllis} For a recent analysis, that allows for
uncertainties in $(m_t,\ m_b)$, and for the relevant references,
see e.g. J. Ellis, K. Olive, Y. Santoso and V. Spanos, Phys. Rev.
{\bf D69}, 095004 (2004), hep-ph/0310356.

\bibitem{phiKs} E. Lunghi, D. Wyler, Phys. Lett. {\bf B521} 320 (2001),
hep-ph/0109149; S. Khalil and E. Kou, Phys. Rev. {\bf D67}, 055009
(2003); hep-ph/0212023; R. Harnik, D. T. Larson, H. Murayama, A.
Pierce, Phys. Rev. {\bf D69},094024 (2004), hep-ph/0212180.


\bibitem{BabarBelleNew}  B. Aubert et al. (BaBar Collaboration), hep-ex/0403026; K. Abe et al. (BELLE Collaboration),
Phys. Rev. Lett. {\bf 91}, 261602 (2003); The most recent results
on $S(B_d\rightarrow \phi K_S)$, submitted to the $32^{nd}$
International Conference of High Energy Physics (Aug. 16-22,
2004), Beijing, China, appear in the papers of B. Aubert et al.
(BaBar Collaboration), hep-ex/0408072, and K. Abe et al. (BELLE
Collaboration), hep-ex/0409049.

\bibitem{Y} See e.g. D. Chang, A. Masiero and H. Murayama, Phys. Rev. {\bf D67},
075013 (2003), hep-ph/0205111; T. Moroi, Phys. Lett. {\bf B493},
366 (2000), hep-ph/0007328; G.L. Kane, P. Ko, H. Wang, C. Kolda,
J. Park, L. Wang, Phys. Rev. {\bf D70}, 035015 (2004),
hep-ph/0212092; J. Park, hep-ph/0312118; J. Hisano and Y. Shimizu,
Phys. Lett. {\bf B565}, 183 (2003), hep-ph/0303071; T. Goto, Y.
Okada, Y. Shimizu, T. Shindou and M. Tanaka, Phys. Rev. {\bf D70},
0305012 (2004), hep-ph/0306093.

\bibitem{Buras2}A. J. Buras, G. Colangelo, G. Isidori, A. Romanino, and L. Silvestrini, Nucl. Phys. {\bf B566},3 (2000); hep-ph/9908371.

\bibitem{Nir} Y. Nir, Lectures Given at Scottish Univ. Summer School, Scotland 2001; hep-ph/0109090.

\bibitem{TBlum} T. Blum et al. (RBC collab.), Phys. Rev. {\bf D68}, 114506
(2003), hep-lat/0110075. See A. Soni, Talk at Pascos 2003
Conference, Mumbai (India), Pramana {\bf 62} 415 (2004),
hep-ph/0307107, for a critical review of this and similar works.

\bibitem{Noaki} J. Noaki et al., Phys. Rev. {\bf D68}, 014501 (2003).

\bibitem{Golterman} M. Golterman and E. Pallante, JHEP {\bf 0110},
037 (2001), Phys. Rev. {\bf D69}, 074503 (2004).

\bibitem{Bhattacharya} T. Bhattacharya, G. T. Fleming, R. Gupta,
G. Kilcup, W. Lee and S. Sharpe, hep-lat/0409046.

\bibitem{Pich} A. Pich, hep-ph/0410215; E. Pallante, A. Pich and
I. Scimemi, Nucl. Phys. {\bf B617}, 441 (2001); E. Pallante and A.
Pich, Phys. Rev. Lett. {\bf 84}, 2568 (2000), Nucl. Phys. {\bf
B592}, 294 (2000).

\bibitem{AAlavi} A. Alavi-Harari et al. (KTev collab.), Phys.
Rev. Lett {\bf 83}, 22 (1999); A. Lai et al. (NA48 collab.), Eur.
Phys. J. {\bf C 22}, 231 (2001).

\bibitem{Nath} T. Ibrahim and P. Nath, Phys. Rev. {\bf D57}, 478
(1998), Phys. Rev. {\bf D58}, 111301 (1998).

\bibitem{AA} G. F. Giudice and A. Masiero, Phys. Lett. {\bf B206}, 480 (1988).

\bibitem{BB} For alternative solutions to the $\mu$-problem based on
gauged $B-L$ symmetry, see: S. King and Q. Shafi, Phys.Lett. {\bf
B422}, 135 (1998);
S. Jeannerot, S. Khalil, G. Lazarides and Q. Shafi, JHEP {\bf
0010}, 012 (2000);
K.~S.~Babu, B.~Dutta and R.~N.~Mohapatra,
Phys.\ Rev.\ D {\bf 65}, 016005 (2002);
R.~Kitano and N.~Okada,
Prog.\ Theor.\ Phys.\  {\bf 106}, 1239 (2001);
L.~J.~Hall, Y.~Nomura and A.~Pierce,
Phys.\ Lett.\ B {\bf 538}, 359 (2002).


\bibitem{C} See e.g. J. Ellis, J. F. Gunion, H. E. Haber, L.
Roszkowski and F. Zwirner, Phys. Rev. {\bf D39}, 844 (1989); L.
Durand and J. L. Lopez, Phys. Lett. {\bf B217}, 463 (1989).

\bibitem{edmneutron} P. Harris et al. Phys. Rev. Lett. {\bf 82},
94 (1999).

\bibitem{edmelectron} E. D. Commins et al, Phys. Rev. {\bf A50},
2960 (1994).

\bibitem{new} For a sample see:
C.~H.~Albright, K.~S.~Babu and S.~M.~Barr,
Phys.\ Rev.\ Lett.\  {\bf 81}, 1167 (1998)
J.~Sato and T.~Yanagida,
Phys.\ Lett.\ B {\bf 430}, 127 (1998)
N.~Irges, S.~Lavignac and P.~Ramond,
Phys.\ Rev.\ D {\bf 58}, 035003 (1998);
C.~H.~Albright and S.~M.~Barr,
Phys.\ Lett.\ B {\bf 461}, 218 (1999);
G.~Altarelli and F.~Feruglio,
Phys.\ Lett.\ B {\bf 451}, 388 (1999); T. Blazek, S. Raby and K.
Tobe, Phys. Rev. {\bf D62}, 055001 (2000),  hep-ph/9912482;
K.~Hagiwara and N.~Okamura,
Nucl.\ Phys.\ B {\bf 548}, 60 (1999);
M.~C.~Chen and K.~T.~Mahanthappa,
Int.\ J.\ Mod.\ Phys.\ A {\bf 18}, 5819 (2003);
C.~S.~Huang, T.~j.~Li and W.~Liao,
Nucl.\ Phys.\ B {\bf 673}, 331 (2003);
N.~Maekawa,
arXiv:hep-ph/0402224.












\end{thebibliography}
\end{document}